\documentclass[english,aps,showpacs,twocolumn,superscriptaddress,floatfix, notitlepage,prl,reprint,unicode=true,colorlinks=true,citecolor=Blue,linkcolor=RubineRed,urlcolor=Blue]{revtex4-1}
\usepackage[T1]{fontenc}
\usepackage[latin9]{inputenc}
\usepackage{color}
\usepackage{babel}
\usepackage{float}
\usepackage{booktabs}
\usepackage{multirow}
\usepackage{amsmath}
\usepackage{amsthm}
\usepackage{amssymb}
\usepackage{graphicx}
\usepackage[unicode=true]
 {hyperref}

\makeatletter

\providecommand{\tabularnewline}{\\}
\floatstyle{ruled}
\newfloat{algorithm}{H}{loa}
\providecommand{\algorithmname}{Algorithm}
\floatname{algorithm}{\protect\algorithmname}

\theoremstyle{plain}
\newtheorem{thm}{\protect\theoremname}

\usepackage{braket}
\usepackage{bm}
\usepackage{txfonts} 
\usepackage{graphicx}
\usepackage[usenames,dvipsnames]{xcolor}
\date{\today}

\makeatother

\providecommand{\theoremname}{Theorem}

\begin{document}
\title{Theory of Compression Channels for Postselected Quantum Metrology}
\author{Jing Yang~\href{https://orcid.org/0000-0002-3588-0832}{\includegraphics[scale=0.05]{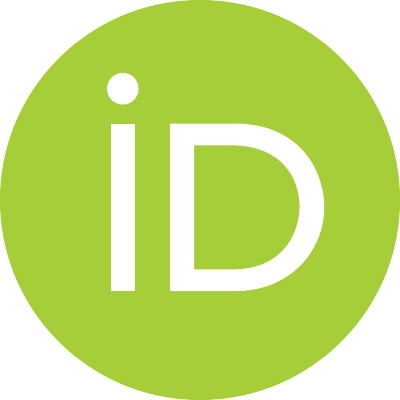}}}
\address{Nordita, KTH Royal Institute of Technology and Stockholm University,
Hannes Alfv\'ens vag 12, 106 91 Stockholm, Sweden}

\begin{abstract}
Postselected quantum
metrological scheme is especially advantageous when the final measurements
are either very noisy or expensive in practical experiments. In this
work, we put forward a general theory on the compression channels
in postselected quantum metrology. We define the basic notions characterizing
the compression quality and illuminate the underlying structure of lossless compression channels. Previous
experiments on Postselected optical phase estimation and weak-value
amplification are shown to be particular cases of this general theory.
Furthermore, for two categories of bipartite systems, we show that
the compression loss can be made arbitrarily small even when the compression
channel acts only on one subsystem. These findings can be employed
to distribute quantum measurements so that the measurement noise and
cost are dramatically reduced. 
\end{abstract}
\maketitle
Quantum metrology utilizes quantum
coherence and entanglement to boost the measurement precision quantified
by the quantum Fisher information (QFI)\,\cite{helstrom1976quantum,holevo2011probabilistic,pezz`e2018quantum,liu2019quantum}. In standard quantum metrology, given an ensemble of metrological samples, quantum mechanics allows one to optimize the quantum measurements so that the information about the signal is maximally extracted. Yet,  another metrological scheme, called Postselected quantum metrology arises in the context of weak-value amplification (WVA)\,\cite{aharonov1988howthe,hosten2008observation,dixon2009ultrasensitive,starling2009optimizing,starling2010precision,aharonov1988howthe,xu2013phaseestimation,feizpour2011amplifying}, where a post-selection measurement is performed to select a sub-ensemble of the samples before the information-extracting measurement. Comparing to standard metrology, though the QFI encoded in the subensemble averaged over its post-selection probability cannot be larger than the QFI in standard metrology\citep{ferrie2014weakvalue,combes2014quantum,tanaka2013information,knee2013quantum,zhu2011quantum,dressel2014colloquium}, there are several advantages due to post-selection {when the cost of the post-selection measurement becomes cheap:} (i) WVA outperforms the standard metrology in the presence of certain types of technical noise~\cite{jordan2014technical}. (ii) WVA can be viewed as a filter to reduce the number of detected samples in standard metrology without losing the precision significantly. As such, in Hamiltonian learning~\citep{anshu2021sampleefficient,haah2022optimal},  post-selection can be employed to reduce   the sample complexity~\cite{jenne2021quantum}, i.e., the number of samples to achieve a given precision. Practically speaking, when the final information-extracting measurements are subjected to various kinds of imperfections, such as detector saturation, limited memory and computational power etc, Postselected quantum metrology is provably outperforms the standard one~\cite{lyons2015powerrecycled,zhang2015precision,harris2017weakvalue,xu2020approaching,chu2020quantum}.

Recently post-selection has been applied in a broad context in quantum
metrology beyond the setup of WVA\,\cite{arvidsson-shukur2020quantum}.
The advantage of post-selection as a filter
or compression channel persists in this broad context as demonstrated
in the experiment of optical phase estimation~\cite{lupu-gladstein2022negative}.
While post-selection can be also applied to classical metrology, previous
works\,\cite{arvidsson-shukur2020quantum,pang2015improving} show
that the non-classicality can further boost the precision. However,
despite these advances, a comprehensive theory for designing
the lossless post-selection measurement channels in the most general
setups beyond WVA remains uncharted. In standard quantum metrology, for arbitrary parameter-dependent quantum states,
the optimal measurements saturating the quantum Cram\'er-Rao
bound were studied by Braunstein and Caves~\cite{braunstein1994statistical}  and recently applied to the case of noisy detection~\cite{len2022quantum}.
Analogously, in Postselected metrology similar fundamental question
has not been addressed. 
\begin{figure}
\begin{centering}
\includegraphics[width=0.7\linewidth]{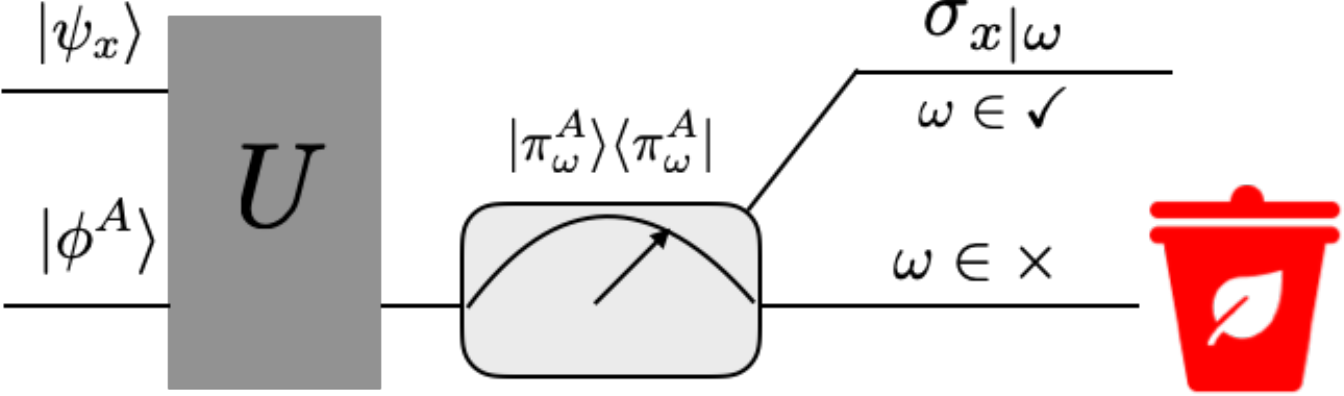}
\par\end{centering}
\caption{\label{fig:Protocol}The protocol of Postselected quantum metrology.
"A" denotes the ancilla. The unitary operation combined with the following projective measurement  on the
ancilla implement the post-selection channel $K_{\omega}$. After post-selection, measurements can be further performed (not shown), as in standard metrology, on these Postselected
states to extract information about the estimation parameter.}
\end{figure}

In this work, we answer this question by proposing a theory that unifies
weak-value metrology, Postselected metrology, as well as standard
metrology. The crucial observation is that standard metrology only
makes use of the measurement statistics and discards the post-measurement
states completely while Postselected metrology utilizes a specific
set of post-measurement states and discards the rest. As such, we
generalize the optimal measurement condition from standard metrology
to Postselected metrology. By keeping track of these conditions,
we identify the generic structure of the lossless post-selection
channel for pure states. Previous setup on Postselected metrology~\cite{arvidsson-shukur2020quantum,lupu-gladstein2022negative,jenne2021quantum,salvati2023compression}
and WVA~\cite{dressel2014colloquium} are special cases of this
general theory. Finally, for bipartite entangled states, when the
compression channel is restricted to one subsystem, we construct two
categories of examples, which can be compressed substantially with
only a negligible amount of loss. 

\textit{Generalized optimal measurement condition. ---} We consider
a pure quantum state of a quantum sensor described by $\rho_{x}\!=\!\ket{\psi_{x}^{\text{}}}\bra{\psi_{x}^{\text{}}}$,
where $x$ is the estimation parameter. The QFI associated with this
state is $I(\rho_{x})\!=\!4g(\rho_{x})\!=\!\sum_{\omega}I_{\omega}(\rho_{x})$, where $g(\rho_{x})\!\equiv\!\braket{\partial_{x}^{\perp}\psi_{x}\big|\partial_{x}^{\perp}\psi_{x}}$, $\ket{\partial_{x}^{\perp}\psi_{x}}\!\equiv\!\ket{\partial_{x}\psi_{x}}\!-\!\braket{\psi_{x}\big|\partial_{x}\psi_{x}}\ket{\psi_{x}}$~\citep{yang2019optimal,zhou2020saturating,braunstein1994statistical}, $I_{\omega}(\rho_{x})\equiv4\braket{\partial_{x}^{\perp}\psi_{x}\big|E_{\omega}\big|\partial_{x}^{\perp}\psi_{x}}$, and $E_{\omega}$ is a positive-operator-valued
measure (POVM) operator, satisfying $E_{\omega}\ge0$ and $\sum_{\omega}E_{\omega}=\mathbb{I}$ ~\citep{nielsen2010quantum}.
In Postselected quantum metrology, a post-selection measurement channel
denoted as $\{K_{\omega}\}$ is performed on the system, where $\omega\in\Omega$
and $\Omega$ is the set of all measurement outcomes. As shown in
Fig.~\ref{fig:Protocol}, such a generalized measurement can be implemented
by a unitary operation entangling the system and ancilla, followed
by a projective measurement on the ancilla. After performing the post-selection
channel, but before post-selection is made, the joint state of the
system and the ancilla becomes $\sigma_{x}^{\text{SA}}\!=\!\sum_{\omega\in\Omega}p(\omega|x)\sigma_{x|\omega}\!\otimes\!\ket{\pi_{\omega}^{A}}\bra{\pi_{\omega}^{A}}$,
where $p(\omega|x)\!=\!\braket{\psi_{x}\big|E_{\omega}\big|\psi_{x}}$
and $\sigma_{x|\omega}\!=\!K_{\omega}\ket{\psi_{x}}\bra{\psi_{x}}K_{\omega}^{\dagger}/p(\omega|x)$~\footnote{We are only concerned with fundamental advantages of post-selections
and therefore do not consider mixed initial states of the ancilla,
which leads to inefficiency in reading the post-selection outcomes,
see Fig.~\ref{fig:Protocol}.}. Throughout this work, states and operators that do not act on the
system only will be specified through the superscript, ``A'', ``SA''
etc. The QFI corresponding to $\sigma_{x}^{\text{SA}}$ is\,\cite{combes2014quantum,SM}, $I(\sigma_{x}^{\text{SA}})\!=\!\sum_{\omega}I_{\omega}(\sigma_{x}^{\text{SA}})$,
where 
\begin{equation}
I_{\omega}(\sigma_{x}^{\text{SA}})\equiv I_{\text{\ensuremath{\omega}}}^{\text{cl}}\left(p(\omega\big|x)\right)+p(\omega\big|x)I(\sigma_{x|\omega}),\label{eq:I-omg}
\end{equation}
and $I_{\text{\ensuremath{\omega}}}^{\text{cl}}\left(p(\omega\big|x)\right)\!\equiv\!\left[\partial_{x}p(\omega\big|x)\right]^{2}\!/\!p(\omega\big|x)$.
\textcolor{black}{The physical meaning of Eq.~(\ref{eq:I-omg}) is
clear: The QFI for each measurement outcome $\omega$ consists of
two parts, the classical QFI and the average QFI for the post-measurement
states. }Clearly, the post-selection channel cannot increase the
QFI, i.e., $I(\sigma_{x}^{\text{SA}})\leq I(\rho_{x}^{\text{}})$.
A more refined statement is the following\,\cite{SM}:
\begin{equation}
I_{\omega}(\sigma_{x}^{\text{SA}})\leq I_{\omega}(\rho_{x}^{\text{}}).\label{eq:meas-ineq-omg}
\end{equation}
If Eq.~(\ref{eq:meas-ineq-omg}) is saturated for all measurement
outcomes, then $I(\sigma_{x}^{\text{SA}})=I(\rho_{x})$. 

To compress the number of samples without sacrificing the precision,
we demand the discarded set contains no information. Therefore, a
precondition to reaching this goal is that Eq.~(\ref{eq:meas-ineq-omg})
must be saturated even before the selection process is made. 
For a regular POVM measurement, where $\braket{\psi_{x}\big|E_{\omega}\big|\psi_{x}}\neq0$
(see \citep{SM} and Ref.~\cite{yang2019optimal} for an elaborated
definition), the necessary and sufficient condition to saturate the
inequality~(\ref{eq:meas-ineq-omg}) is given by Eq.~(\ref{Teq:Im-perp-E})
in Table~\ref{tab:saturation-cond}. For a null POVM measurement
where $\braket{\psi_{x}\big|E_{\omega}\big|\psi_{x}}\!=\!0$, $I_{\text{\ensuremath{\omega}}}^{\text{cl}}\left(p(\omega\big|x)\right)\!=\!I_{\omega}(\rho_{x}^{\text{}})$
and $I(\sigma_{x|\omega})\!=\!0$, see Ref.~\citep{yang2019optimal}
for details. Thus no information is left in the post-measurement state
$\sigma_{x|\omega}$. As a consequence, if $\sigma_{x|\omega}$ is
the state one would like to retain, one should avoid designing $E_{\omega}$
as a null POVM measurement operator. 

In standard quantum metrology, the post-measurement states are all
discarded and only the measurement statistics is retained. In this
case, one would like to saturate the inequality, $I_{\omega}^{\text{cl}}\left(p(\omega\big|x)\right)\!\leq\! I_{\omega}(\rho_{x}^{\text{}})$,
which was studied by the classic work of Braunstein and Caves~\citep{braunstein1994statistical} and the recent work Ref.~\cite{len2022quantum},
see also Eq.~(\ref{Teq:prop-cond-r}). In Postselected metrology,
we require the average QFI of the retained post-measurement state
saturates the quantum limit, i.e., $p(\omega\big|x)I(\sigma_{x|\omega}^{\text{}})\!\leq\! I_{\omega}(\rho_{x}^{\text{}})$.
Finally, for the measurement outcome or the corresponding post-measurement
state to be discarded, we would like them to carry no information,
i.e., $I_{\text{\ensuremath{\omega}}}^{\text{cl}}\left(p(\omega\big|x)\right)\!=\!0$
or $I(\sigma_{x|\omega}^{\text{}})\!=\!0$. The saturation conditions
of these bounds are given in Table~\ref{tab:saturation-cond}. We
shall call Eqs.~(\ref{Teq:Im-perp-E}-\ref{Teq:prop-cond-r}) as
the \textit{generalized} optimal measurement conditions, as they generalize
the results by Braunstein and Caves~\cite{braunstein1994statistical}
to account for the case where the information about the parameter
is losslessly encoded either in the measurement statistics or the
post-measurement states. As we will see, they play a fundamental role
in the theory of compression channels. 

\newcounter{tableeqn}[table] 
\renewcommand{\thetableeqn}{T\arabic{tableeqn}}

\begin{table}
\caption{\label{tab:saturation-cond}The necessary and sufficient conditions
for the saturation of the bounds of various QFIs corresponding to
a regular POVM operator $E_{\omega}$ with $\sqrt{E_{\omega}}\ket{\psi_{x}}\!\neq\!0$. See Supplemental Material for details~\citep{SM}.}

\begin{centering}
\begin{tabular}{ccc}
\toprule 
Saturation & \multicolumn{2}{c}{Necessary and sufficient condition}\tabularnewline
\midrule 
$I_{\omega}(\sigma_{x}^{\text{SA}})=I_{\omega}(\rho_{x}^{\text{}}),$ & $\ensuremath{{\displaystyle \text{Im}\braket{\partial_{x}^{\perp}\psi_{x}\big|E_{\omega}\big|\psi_{x}}=0.}}$ & \refstepcounter{tableeqn} (\thetableeqn)\label{Teq:Im-perp-E}\tabularnewline
\midrule
\multirow{1}{*}{$I(\sigma_{x|\omega}^{\text{}})=0,$} & $\sqrt{E_{\omega}}\ket{\partial_{x}^{\perp}\psi_{x}}=c\sqrt{E_{\omega}}\ket{\psi_{x}},\,c\in\mathbb{C}$. & \refstepcounter{tableeqn} (\thetableeqn)\label{Teq:prop-cond-c}\tabularnewline
$I_{\text{\ensuremath{\omega}}}^{\text{cl}}\left(p(\omega\big|x)\right)=0,$ & $\text{Re}\braket{\partial_{x}^{\perp}\psi_{x}\big|E_{\omega}\big|\psi_{x}}=0$. & \refstepcounter{tableeqn} (\thetableeqn)\label{Teq:Re-vanish}\tabularnewline
\midrule
$p(\omega\big|x)I(\sigma_{x|\omega}^{\text{}})=I_{\omega}(\rho_{x}^{\text{}}),$ & $\sqrt{E_{\omega}}\ket{\partial_{x}^{\perp}\psi_{x}}\perp\sqrt{E_{\omega}}\ket{\psi_{x}}.$ & \refstepcounter{tableeqn} (\thetableeqn)\label{Teq:perp-cond}\tabularnewline
$I_{\text{\ensuremath{\omega}}}^{\text{cl}}\left(p(\omega\big|x)\right)=I_{\omega}(\rho_{x}^{\text{}}),$ & $\sqrt{E_{\omega}}\ket{\partial_{x}^{\perp}\psi_{x}}=c\sqrt{E_{\omega}}\ket{\psi_{x}},\,c\in\mathbb{R}.$ & \refstepcounter{tableeqn} (\thetableeqn)\label{Teq:prop-cond-r}\tabularnewline
\bottomrule
\end{tabular}
\par\end{centering}
\end{table}

\textit{Lossless Compression Channel.--- }In the standard metrology, the discarded set, as indicated by the
red trash bin in Fig.~\ref{fig:Protocol}, is \includegraphics[scale=0.15]{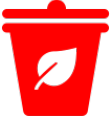}$\!=\!\{\sigma_{x|\omega}\}_{\omega\in\Omega}$
and no QFI is left in the post-measurement states. In Postselected
metrology, we require all the QFI is transferred to the desired post-measurement
states and the discarded set contains no information. 
We use $\omega\in\checkmark$ and $\omega\in\times$
to indicate the desired and undesired outcomes, respectively. Throughout
this work, we consider a minimum retained set $\{\sigma_{x|\omega}\}_{\omega\in\checkmark}$,
corresponding to \includegraphics[scale=0.15]{TrashBin.pdf}$=\Omega\cup\{\sigma_{x|\omega}\}_{\omega\in\times}$
 \footnote{Alternatively, one may also wish to retain the desired outcomes,
where the discarded set is $\{\sigma_{x|\omega}\}_{\omega\in\times}$,
which will be discussed elsewhere. With Table~\ref{tab:saturation-cond},
the corresponding theory for the latter can be developed analogously. }. 
When $\sum_{\omega\in\checkmark}p(\omega|x)<1$, we can view the post-selection
as a compression channel. It is worth noting that even if $\sum_{\omega\in\checkmark}p(\omega|x)$ is small, resulting in a small number of metrological samples in the Postselected ensemble in each round,  the experiment is assumed to be repeated for a sufficiently large number of rounds so that the classical Cram\'er-Rao bound is saturated.

Let us first introduce the essential notions for the theory of compression
channels. The loss of the QFI per input sample can be expressed as
$\gamma\!\equiv\!1-\sum_{\omega\in\checkmark}p(\omega|x)I_{\omega}(\sigma_{x|\omega})/I(\rho_{x})$.
We define $c\equiv1/\sum_{\omega\in\checkmark}p(\omega|x)$ as the\textit{
compression capacity} for a post-selection channel, {characterizing the ability of a designed post-selection measurement to reduce the number of samples}. If $\gamma=0$ and $c>1$ then a post-selection measurement
is called an lossless compression channel (LCC). We shall restrict our attention to \textit{efficient}
post-selection channels where $c\in(1,\,\infty)$~\footnote{As mentioned previously, for $\omega\in\checkmark$, $E_{\omega}$
should not be null POVM operator and therefore $p(\omega|x)$ must be strictly positive and $c$ cannot blow up. One may think in the
limit $c=\infty$ and $p(\omega\in\checkmark|x)=0$, the compression is too strong so that the sample ``crushes'' and we will not collect any desired samples.}. We further define
the \textit{compression gain}, $\eta\equiv\sum_{\omega\in\checkmark}I_{\omega}(\sigma_{x|\omega}^{\text{}})/I(\rho_{x}^{\text{}})$,
as the ratio between the  Postselected QFI
and the one standard metrology, {characterizing information gain per detected sample.} This characterizes the advantage of the former
over the latter when the cost of final detection dominates over the cost of post-selection~\citep{arvidsson-shukur2020quantum}. 

For generic quantum systems, we find LCC must satisfy the following
theorem:
\begin{thm}
\label{thm: LCC-general}\textup{For a pure state $\ket{\psi_{x}}$, the POVM
operators in an efficient LCC must satisfy
\begin{align}
\braket{\psi_{x}^{\perp}\big|\sum_{\omega\in\checkmark}E_{\omega}\big|\psi_{x}^{\perp}} & =1,\label{eq:E-omg-cross}\\
\braket{\partial_{x}^{\perp}\psi_{x}\big|E_{\omega}\big|\psi_{x}} & =0,\label{eq:E-checkmark-general}
\end{align}
with $p(\omega|x)>0$ for $\omega\in\checkmark$ and $\sum_{\omega\in\checkmark}p(\omega|x)<1$,
where $\ket{\psi_{x}^{\perp}}\equiv\ket{\partial_{x}^{\perp}\psi_{x}}/\sqrt{g(\rho_{x})}$
is the normalized vector along $\ket{\partial_{x}^{\perp}\psi_{x}}$
direction. }
\end{thm}
The proof is straightforward with the following intuition: Eq.~(\ref{eq:E-omg-cross})
guarantees that in the undesired outcome,
measurements statistics and post-measurement states contain no QFI,
i.e., $I_{\omega}(\rho_{x}^{\text{}})=0$ for  $\omega\in\times$; Eq.~(\ref{eq:E-checkmark-general})
ensures that for the desired outcome where $\omega\in\checkmark$,
the measurements statistics again contains no QFI and retained states
reach the quantum limit given by Eq.~(\ref{Teq:perp-cond}). As such,
the QFI is fully preserved after the post-selection. Alternatively,
the following theorem illuminates the underlying structure of an
LCC:
\begin{thm}
\label{thm: LCC-structure} \textup{For a pure state $\ket{\psi_{x}}$, the
POVM operators in the retained set of an LCC can be expressed
as follows:
\begin{equation}
E_{\omega}=q_{\omega}\rho_{x}^{\perp}+\Lambda_{\omega},\label{eq:E-omg-checkmark}
\end{equation}
where $q_{\omega}\in(0,1]$ satisfying $\sum_{\omega \in \checkmark} q_{\omega}=1$,
$\rho_{x}^{\perp}\equiv\ket{\psi_{x}^{\perp}}\bra{\psi_{x}^{\perp}}$
and $\Lambda_{\omega}$ is the gauge operator, which does not contributes
to the QFI and can be chosen in many ways as long as it satisfies
\begin{equation}
\braket{\psi_{x}^{\perp}\big|\Lambda_{\omega}\big|\psi_{x}^{\perp}}=\braket{\psi_{x}^{\perp}\big|\Lambda_{\omega}\big|\psi_{x}}=0,\label{eq:Lambda-Condition}
\end{equation}
and $\,\braket{\psi_{x}\big|\Lambda_{\omega}\big|\psi_{x}}=\lambda_{\omega}\in(0,\,1)$.
The compression capacity and gain are $c=1/\sum_{\omega\in\checkmark}\lambda_{\omega}$
and $\eta=\sum_{\omega\in\checkmark}q_{\omega}/\lambda_{\omega}$,
respectively.}
\end{thm}
Theorem~\ref{thm: LCC-general} and \ref{thm: LCC-structure} are
our second main results. Practically, the LCC~(\ref{eq:E-omg-checkmark})
depends on the true value of $x$, so in general adaptive estimation
is required~\citep{demkowicz-dobrzanski2014usingentanglement,yuan2015optimal,pang2017optimal, schmitt2017submillihertz, gefen2017control, yang2017quantum,demkowicz-dobrzanski2017adaptive}.
If we assume full accessibility of the post-selection measurements
on the whole Hilbert space, i.e., Eq.~(\ref{eq:E-omg-checkmark})
is always implementable regardless of the choice of $\Lambda_{\omega}$,
then $\lambda_{\omega}$ can be tuned arbitrarily small, say $\lambda_{\omega}=\varepsilon$.
Then we find $\eta=Lc=1/\varepsilon$, where $L$ is the number of
desired outcomes. 

When $\eta>1$, we know the QFI per detected sample in an LCC is amplified,
though it will be counter-balanced when the post-selection probability
is accounted for~\citep{arvidsson-shukur2020quantum,combes2014quantum,ferrie2014weakvalue}.
While Ref.~\citep{arvidsson-shukur2020quantum} relates such an amplification
to the non-commutativity of observables, thanks to Theorem~\ref{thm: LCC-general}
and \ref{thm: LCC-structure}~, we can directly compute the enhancement
of the parametric sensitivity of the Postselected state to the estimation
parameter. For example, apart from an irrelevant unitary rotation,
one can take $K_{\omega}=\sqrt{q_{\omega}}\rho_{x}^{\perp}+\sqrt{\lambda_{\omega}}\rho_{x}$,
corresponding to $\Lambda_{\omega}=\lambda_{\omega}\rho_{x}$. While
the post-measurement state is still $\ket{\psi_{x}}$, the parameter
derivative of the Postselected state becomes~\citep{SM}:
\begin{equation}
\partial_{x}\left(K_{\omega}\ket{\psi_{x}}/\sqrt{p(\omega|x)}\right)=\ket{\partial_{x}\psi_{x}}+\left(\sqrt{q_{\omega}/\lambda_{\omega}}-1\right)\sqrt{g(\rho_{x})}\ket{\psi_{x}^{\perp}}.
\end{equation}

Finally, it is worth to note that if $\lambda_{\omega}=0$, the LCC~(\ref{eq:E-omg-checkmark})
degenerates into a null measurement operator $q_{\omega}\rho_{x}^{\perp}$.
Using our prior knowledge about the estimation parameter denoted as
$x_{*}$, the projector $\rho_{x_{*}}^{\perp}$ can approach to
quantum limit asymptotically as $x_{*}\to x$~\citep{yang2019optimal}. 

\textit{Binary post-selection}.\textit{---} For a binary post-selection,
i.e., the desired set $\checkmark$ contains only one single outcome.
In this case, for simplicity we use $\checkmark$ as a shorthand
notation for $\omega\in\checkmark$. Apparently $q_{\checkmark}=1$
and $c=\eta=1/\lambda_{\checkmark}$, but there are many choices of
choosing $\Lambda_{\checkmark}$. If we take the gauge operator $\Lambda_{\checkmark}=\lambda_{\checkmark}\rho_{x}+\mathcal{P}_{0}$,
where $\mathcal{P}_{0}$ is the projector to the orthogonal complement
to the subspace spanned by $\ket{\psi_{x}}$ and $\ket{\psi_{x}^{\perp}}$,
then Eq.~(\ref{eq:E-omg-checkmark}) becomes $E_{\checkmark}=(\lambda_{\checkmark}-1)\rho_{x}+\mathbb{I}$,
which is the LCC proposed in Refs.~\citep{jenne2021quantum,salvati2023compression}. 

In two-level systems, the gauge operator $\Lambda_{\checkmark}$ is
forced to take the form $\Lambda_{\checkmark}=\lambda_{\checkmark}\rho_{x}$.
Consider a parameter-dependent state $\ket{\psi_{x}}=\cos\left(\frac{x\Delta}{2}\right)\ket{0}+\text{i}\sin\left(\frac{x\Delta}{2}\right)\ket{1}$,
where $\Delta$ is a known constant and $\{\ket{0}$, $\ket{1}\}$
is a parameter-independent basis. This is the example investigated
in the experiment in Ref.~\citep{lupu-gladstein2022negative}. Our
theory predicts that the Postselected POVM measurement operator in
the basis of $\{\ket{0},\,\ket{1}\}$ is 
\begin{equation}
E_{\checkmark}=\begin{bmatrix}\lambda_{\checkmark}\cos^{2}\left(\frac{x\Delta}{2}\right)+\sin^{2}\left(\frac{x\Delta}{2}\right) & \text{i}(1-\lambda_{\checkmark})\sin(x\Delta)\\
-\text{i}(1-\lambda_{\checkmark})\sin(x\Delta) & \cos^{2}\left(\frac{x\Delta}{2}\right)+\lambda_{\checkmark}\sin^{2}\left(\frac{x\Delta}{2}\right)
\end{bmatrix}.\label{eq:E-cm-2-level}
\end{equation}
It should be noted that unlike Ref.~\citep{lupu-gladstein2022negative}
which assumes small values of $x\Delta$, Eq.~(\ref{eq:E-cm-2-level})
is the exact LCC for any values of $x$. Of course, when $x\Delta<<1$,
to the zeroth order of $x\Delta$, it recovers the LCC in Ref.~\citep{lupu-gladstein2022negative},
i.e., $E_{\checkmark}=\lambda_{\checkmark}\ket{0}\bra{0}+\ket{1}\bra{1}$.

\textit{Restricted post-selections.--- }When $\ket{\psi_{x}}$ is
a bipartite entangled state between two subsystems $A$ and $B$,
the post-selection measurement on only $A$ generically leads to loss
i.e. $\gamma>0$. This is because the LCC, according to Theorem~\ref{thm: LCC-structure},
in general acts globally on both the system and the environment.
Nevertheless, we demonstrate the existence of approximate LCC for
two classes of examples, where the loss is very tiny. To
proceed, let us define $\mathcal{C}_{x}^{\text{AB}}\!\equiv\!\ket{\partial_{x}^{\perp}\psi_{x}^{\text{AB}}}\bra{\psi_{x}^{\text{AB}}},\,\mathcal{C}_{x}^{\text{A}}\!=\!\text{Tr}_{\text{B}}\mathcal{C}_{x}^{\text{AB}},\,\varrho_{x}^{\perp\text{A}}\!=\!\text{Tr}_{\text{B}}\rho_{x}^{\perp\text{AB}},\:\varrho_{x}^{\text{A}}=\text{Tr}_{\text{B}}\rho_{x}^{\text{AB}}.$
Consider a post-selection channel on the subsystem $A$ only, i.e.,
$E_{\omega}^{(\text{A})}\otimes\mathbb{I}^{(\text{B})}$. Then Theorem~\ref{thm: LCC-general}
becomes $\text{\text{Tr}\ensuremath{\left(\varrho_{x}^{\perp\text{A}}\sum_{\omega\in\checkmark}E_{\omega}^{\text{A}}\right)}}\!=\!1$, and $\text{Tr}\left(\mathcal{C}_{x}^{\text{A}}E_{\omega}^{\text{A}}\right)\!=\!0,\,\omega\in\checkmark$,
with $\text{Tr}\left(\varrho_{x}^{\text{A}}\sum_{\omega\in\checkmark}E_{\omega}^{\text{A}}\right)\!<\!1$. 

The first category of examples is the weak-entanglement limit, which
includes WVA as a special case. We consider a separable pure initial
state $\ket{\psi_{0}^{\text{AB}}}\equiv\ket{\phi_{0}^{\text{A}}}\otimes\ket{\varphi_{0}^{\text{B}}}$.
The Hamiltonian is $H_{\text{AB}}\!=\!x(H_{\text{A}}\otimes H_{\text{B}})$.
By a judicious choice of the initial states, one can always make $\braket{\phi_{0}^{\text{A}}\big|H_{\text{A}}\big|\phi_{0}^{\text{A}}}=0$
or $\braket{\varphi_{0}^{\text{B}}\big|H_{\text{B}}\big|\varphi_{0}^{\text{B}}}=0$
such that $\braket{\psi_{0}^{\text{AB}}\big|H_{\text{AB}}\big|\psi_{0}^{\text{AB}}}=0$.
The QFI is $I(\rho_{x})=4\left(\braket{\phi_{0}^{\text{A}}\big|H_{\text{A}}^{2}\big|\phi_{0}^{\text{A}}}\braket{\varphi_{0}^{\text{B}}\big|H_{\text{B}}^{2}\big|\varphi_{0}^{\text{B}}}\right).$
In local estimation theory, $x$ is usually considered to be very
small. In the limit $x\to0$, we find $\ket{\psi_{x}}\!=\!\ket{\phi_{0}^{\text{A}}}\otimes\ket{\varphi_{0}^{\text{B}}}$
and $\ket{\partial_{x}^{\perp}\psi_{x}}\!=\!-\text{i}H_{\text{A}}\ket{\phi_{0}^{\text{A}}}\otimes H_{\text{B}}\ket{\varphi_{0}^{\text{B}}}$
are disentangled. Then it can be readily calculated that $\varrho_{x}^{\perp A}\!=\!H_{\text{A}}\ket{\phi_{0}^{\text{A}}}\bra{\phi_{0}^{\text{A}}}H_{\text{A}}/\braket{\phi_{0}^{\text{A}}\big|H_{\text{A}}^{2}\big|\phi_{0}^{\text{A}}}$,
$\varrho_{x}^{A}\!=\!\ket{\phi_{0}^{\text{A}}}\bra{\phi_{0}^{\text{A}}}$,
and $\mathcal{C}_{x}^{\text{A}}=H_{\text{A}}\ket{\phi_{0}^{\text{A}}}\bra{\phi_{0}^{\text{A}}}$.

Now if $\braket{\phi_{0}^{\text{A}}\big|H_{\text{A}}\big|\phi_{0}^{\text{A}}}\!=\!0$,
similar with Eq.~(\ref{eq:E-omg-checkmark}), one can construct
\begin{equation}
E_{\omega}^{\text{A}}=q_{\omega}\varrho_{x}^{\perp A}+\varepsilon\ket{\phi_{0}^{\text{A}}}\bra{\phi_{0}^{\text{A}}},\label{eq:E-singular}
\end{equation}
where $\sum_{\omega\in\checkmark}q_{\omega}\!=\!1$, $\varepsilon$ is
arbitrarily small. In this case $\eta\!=\!Lc\!=\!1/\varepsilon$ as before,
i.e., we can achieve arbitrarily large compression capacity and gain
without loss. On the other hand, if $\braket{\phi_{0}^{\text{A}}\big|H_{\text{A}}\big|\phi_{0}^{\text{A}}}\neq0$
but $\braket{\varphi_{0}^{\text{B}}\big|H_{\text{B}}\big|\varphi_{0}^{\text{B}}}=0$.
One can simply take 
\begin{equation}
E_{\omega}^{\text{A}}=q_{\omega}\varrho_{x}^{\perp A}.\label{eq:E-regular}
\end{equation}
In this case, the compression capacity is $c\!=\!\braket{\phi_{0}^{\text{A}}\big|H_{\text{A}}^{2}\big|\phi_{0}^{\text{A}}}/\braket{\phi_{0}^{\text{A}}\big|H_{\text{A}}\big|\phi_{0}^{\text{A}}}^{2}$,
which is the ratio between the second and first order moments of
the energy of the subsystem $A$. The compression gain is $\eta\!=\!Lc$. 

It is worth noting that WVA~\citep{dressel2014colloquium,aharonov1988howthe}
falls into this category. Considering the von-Neumann measurement
model~\citep{vonneumann1955mathematical}, where the system consists
of a two-level subsystem and the continuous-variable meter. The Hamiltonian
of the system is $H\!=\!x\sigma_{z}\!\otimes\! P_{u}$,
where $P_{u}\!=\!-\text{i}\partial\!/\!\partial u$ and $u$ is the position
of the meter. The initial state is $\ket{\phi_{\theta}}\otimes\ket{\varphi_{0}^{\text{}}},$where
$\ket{\phi_{\theta}}\!=\!\cos\left(\frac{\theta}{2}\right)\ket{0}\!+\!\sin\left(\frac{\theta}{2}\right)\ket{1}$
and $\ket{\varphi_{n}}\!=\!\int du\varphi_{n}(u)\ket{u}$ with $\varphi_{n}(u)$
is the $n$-th order normalized Hermite-Gaussian function defined as $\varphi_{n}(u)=\frac{1}{\sqrt{2^{n}n!}}\left(\frac{1}{2\pi\sigma^{2}}\right)^{1/4}e^{-u^{2}/(4\sigma^{2})}H_{n}\left(\frac{u}{\sqrt{2}\sigma}\right)$, where $H_n(x)$ is well-known Hermite polynominal~\cite{sakurai2010modernquantum}. The parameter-dependent
state is $\ket{\psi_{x}^{\text{}}}\!=\!e^{-\text{i}x\sigma_{z}\otimes\hat{P}}\ket{\phi^{\text{}}}\otimes\ket{\varphi_{0}^{\text{}}}$.
The QFI before post-selection is $I(\rho_{x})\!=\!1/\sigma^{2}$. One
can readily show $\braket{\varphi_{0}\big|P_{u}\big|\varphi_{0}}\!=\!0$.
In the limit $x\!\to\!0$, one can find $\ket{\psi_{x\!=\!0}}\!=\!\ket{\phi_{\theta}}\otimes\ket{\varphi_{0}}$
and $\ket{\psi_{x\!=\!0}^{\perp}}\!=\!\ket{\phi_{-\theta}}\!\otimes\!\ket{\varphi_{1}}$.
The WVA employs a binary post-selection channel $E_{\checkmark}^{\text{WVA}}\!=\!\ket{\phi_{\theta_{*}}}\bra{\phi_{\theta_{*}}}\otimes\mathbb{I}$, where $\ket{\phi_{\theta_{*}}}$ is almost orthogonal to the initial
spin state $\ket{\phi_{\theta}}$, i.e., $\theta_{*}\!=\!\theta-\pi+2\varepsilon$
and $\varepsilon$ is an arbitrarily small but nonzero real number.
Apparently, the WVA post-selection channel satisfies Eq.~(\ref{eq:E-checkmark-general}),
meaning the measurement statistics associated with the retained set
contains no QFI. However, the loss $\gamma\!=\!\cos^{2}(\theta+\varepsilon)$,
indicating there are information loss in the undesired measurement
statistics and undesired post-measurement states, unless $\theta\!=\!\pi/2$,
where the loss is $\sin^{2}\varepsilon$. 

\begin{figure}
\begin{centering}
\includegraphics[scale=0.43]{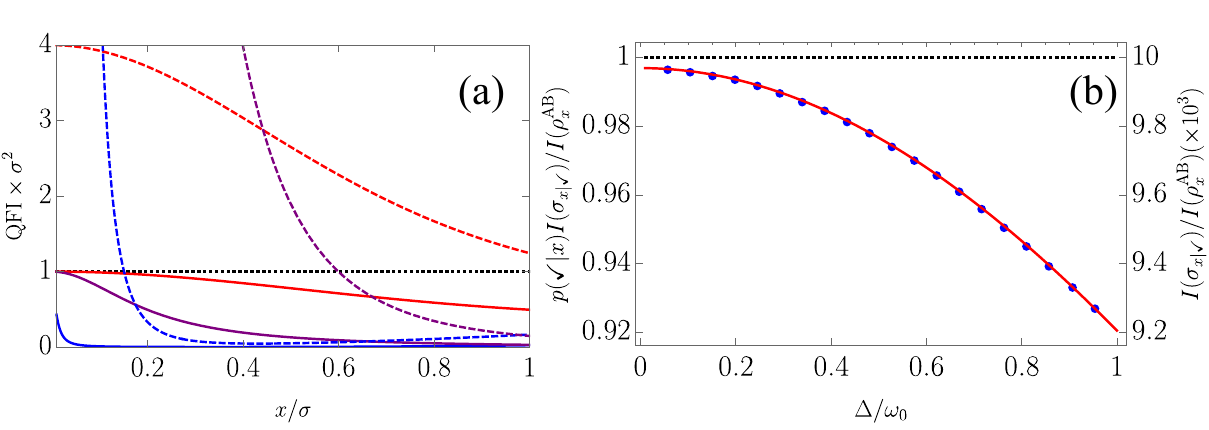}
\par\end{centering}
\caption{\label{fig:num-QFI}In both figures, $K_{\checkmark}\!=\!\sqrt{E_{\checkmark}}$.
(a) The solid lines correspond to $p(\checkmark|x)I(\sigma_{x|\checkmark})/I(\rho_{x})$
with $I(\rho_{x})\!=\!1/\sigma^{2}$, where the deviation from $1$ represents
the loss $\gamma$, while the dashed lines correspond to the gain
$\eta=I(\sigma_{x|\checkmark})/I(\rho_{x})$. Each color represents
one post-selection scheme: Red-LCC on the two-level system
with $\theta_{*}\!=\!-\theta\!=\!-\pi/3$; Blue-WVA with $\theta_{*}\!=\!-2\pi/3\!+\!10^{-2}$;
Purple-LCC on the meter discussed in the main text with $\varepsilon\!=\!10^{-4}$.(b)
Post-selection on the three-qubit entangled state. The LCC is given
by Eq.~(\ref{eq:E-entangled}) with $x=10^{-5}$, $\theta\!=\!\pi/3$, $p_{1}=1\!-\!p_{2}\!=\!2/3$,
and $\varepsilon\!=\!10^{-4}$. The red solid line corresponds to the
plot of $1-\gamma$ (left frame ticks)while the blue round dots correspond
to the gain $\eta$ (right frame ticks). }
\end{figure}

On the other hand, Eqs.~(\ref{eq:E-singular},~\ref{eq:E-regular})
predict that the LCC acting on the two-level system is of the same
form as the WVA post-selection channel, but with a different choice of $\theta_{*}$.
Here $\theta_{*}\!=\!-\theta$ if $\theta\neq\pi/2$ and $\theta_{*}\!=\!-\theta+2\varepsilon$
if $\theta\!=\!\pi/2$. The compression gain is $\eta\!=\!c\!=\!1/\cos^{2}\theta$
if $\theta\!\neq\!\pi/2$ and $1/\sin^{2}\varepsilon$ if $\theta=\pi/2$,
which is consistent with previous analysis. Furthermore, the same
LCC can be alternatively constructed by a judicious choice of the
gauge operator $\Lambda_{\checkmark}$ so that Eq.~(\ref{eq:E-omg-checkmark})
becomes a one-body operator that only acts on the two-level systems~\citep{SM}.
If we consider post-selecting on the meter, Eq.~\eqref{eq:E-singular}
predicts that  $E_{\checkmark}\!=\!\mathbb{I}\otimes\left(\ket{\varphi_{1}}\bra{\varphi_{1}}\!+\!\varepsilon\ket{\varphi_{0}}\bra{\varphi_{0}}\right)$ is also an LCC with $\eta=c=1/\varepsilon^{2}$. Interestingly, measurement
of this type with $\varepsilon=0$ was previously explored to reach
the super-resolution of incoherent imaging~\citep{tsang2016quantum,zhou2019quantumlimited}.
The performance of these LCCs is numerically calculated in Fig.~\ref{fig:num-QFI}.

Another category of examples with negligible loss is when the energy
fluctuation of the Postselected subsystems dominates over the other
if they are non-interacting but the initial state is entangled. We
consider the Hamiltonian of the two systems is $H_{\text{AB}}\!=\!x(H_{\text{A}}\!+\!H_{\text{B}})$.
Denote the eigenstates of $H_{\text{A}}$ as $H_{\text{A}}\ket{E_{n}^{\text{A}}}\!=\!E_{n}^{\text{A}}\ket{E_{n}^{\text{A}}}$,
where $n\!=\!1,\,2\cdots,\,d_{\text{A}}\!\equiv\!\dim\mathcal{H}_{\text{A}}$
and $E_{1}^{\text{A}}\!\leq\! E_{2}^{\text{A}}\!\leq\!\cdots\!\leq\! E_{d_{\text{A}}}^{\text{A}}$.
We then split the Hilbert space into several orthogonal and disjoint
subspaces spanned by the energy eigenstates, i.e., $\mathcal{H}_{\text{A}}\!=\!\oplus_{k}\mathcal{V}_{k}^{\text{A}}$
with $\mathcal{V}_{k}^{\text{A}}\cap\mathcal{V}_{l}^{\text{A}}\!=\!\{0\}$
for $k\neq l$ so that one can construct a set of mutually orthogonal
states with the same average energy, i.e., $\braket{\phi_{k}^{\text{A}}\big|H_{\text{A}}\big|\phi_{k}^{\text{A}}}\!=\!\mathcal{E}$
with $\ket{\phi_{k}^{\text{A}}}\in\mathcal{V}_{k}^{\text{A}}$ is
a superposition of energy eigenstates to ensure non-vanishing QFI.

We consider entangled initial state $\ket{\psi_{0}^{\text{AB}}}\!=\!\sum_{k}\sqrt{p_{k}}\ket{\phi_{k}^{\text{A}}}\!\otimes\!\ket{\varphi_{k}^{\text{B}}}$, where $p_{k}>0$ satisfying $\sum_k p_k\!=\!1$, $\{\ket{\varphi_{k}^{\text{B}}}\}$ is a set of orthonormal
basis on the subsystems $B$. The QFI before post-selection is $I(\rho_{x}^{\text{AB}})\!=\!4(\delta h_{\text{A}}^{2}\!+\!\delta h_{\text{B}}^{2})\approx4\delta h_{\text{A}}^{2}$,
where $\delta h_{\text{A},\,\text{B}}\equiv\sqrt{\text{Var}(H_{\text{A},\,\text{B}})_{\varrho_{0}^{\text{A},\,\text{B}}}}$
and $\delta h_{\text{B}}\ll\delta h_{\text{A}}$ is assumed. While an LCC exists for arbitrary
value of $x$~\citep{SM},  as before,
we focus on the local estimation for small $x$ and consider the following LCC
\begin{equation}
E_{\omega}^{A}=\sum_{k}r_{\omega k}\ket{\phi_{k}^{\perp\text{A}}}\bra{\phi_{k}^{\perp\text{A}}}+\varepsilon\mathcal{P}_{\text{supp}(\varrho_{0}^{\text{A}})},\label{eq:E-entangled}
\end{equation}
where $\varepsilon$ is an arbitrarily small positive number, $\sum_{\omega\in\checkmark}r_{\omega k}=1$, $\ket{\phi_{k}^{\perp\text{A}}}\equiv(H_{\text{A}}-\mathcal{E})\ket{\phi_{k}^{\text{A}}}/\sqrt{\text{Var}(H_{\text{A}})_{\ket{\phi_{k}^{\text{A}}}}}$
satisfying $\braket{\phi_{k}^{\perp\text{A}}\big|\phi_{l}^{\text{A}}}\!=\!0$
and $\braket{\phi_{k}^{\perp\text{A}}\big|\phi_{l}^{\perp\text{A}}}\!=\!\delta_{kl}$
and $\mathcal{P}_{\text{supp}(\varrho^{\text{A}})}$ is any projector
to the support of the reduced density matrix $\varrho_{0}^{\text{A}}\!=\!\sum_{k}p_{k}\ket{\phi_{k}^{\text{A}}}\bra{\phi_{k}^{\text{A}}}$.
Eq.~(\ref{eq:E-entangled}) generalizes Eq.~(\ref{eq:E-omg-checkmark})
beautifully while preserving the similar structure. The scaling of
loss, capacity and gain in this case are $\gamma\sim O[(\delta h_{\text{B}}/\delta h_{\text{A}})^{2}]\!+\!O(L\varepsilon)$,
$c\!\sim\!1/O(L\varepsilon)$, and $\eta\!\sim\!\{1\!-\!O[(\delta h_{\text{B}}/\delta h_{\text{A}})^{2}]\}/O(\varepsilon)$.

For example, consider $A$ and $B$ consists of two qubits and one
qubit respectively with the Hamiltonian $H_{\text{A}}\!=\!\omega_{0}(\sigma_{z}^{(1)}+\sigma_{z}^{(2)})$
and $H_{\text{B}}\!=\!\Delta\sigma_{z}^{(3)}$. The initial state is $\ket{\psi_{0}^{\text{AB}}}\!=\!\sqrt{p_{1}}\ket{\phi_{1}^{\text{A}}}\otimes\ket{\varphi_{1}^{\text{B}}}\!+\!\sqrt{p_{2}}\ket{\phi_{2}^{\text{A}}}\otimes\ket{\varphi_{2}^{\text{B}}}$,
where $\ket{\phi_{1}^{\text{A}}}\!=\!(\ket{00}\!+\!\ket{11})/\sqrt{2}$ ,
$\ket{\phi_{2}^{\text{A}}}\!=\!(\ket{01}\!+\!\ket{10})/\sqrt{2}$, $\ket{\varphi_{1}^{\text{B}}}\!=\!\ket{\phi_{\theta}}$ and
$\ket{\varphi_{1}^{\text{B}}}\!=\!\ket{\phi_{\theta-\pi}}$, where $\ket{\phi_{\theta}}$
is defined previously. We consider binary post-selection and employ
the LCC $E_{\checkmark}^{\text{A}}\!=\!\ket{\phi_{1}^{\perp\text{A}}}\bra{\phi_{1}^{\perp\text{A}}}\!+\!\varepsilon\ket{\phi_{2}^{\text{A}}}\bra{\phi_{2}^{\text{A}}}$,
where $\ket{\phi_{1}^{\perp\text{A}}}\!=\!(\ket{00}\!-\!\ket{11})\!/\!\sqrt{2}$
is orthogonal to $\ket{\phi_{1}^{\text{A}}}$ and $\ket{\phi_{2}^{\perp\text{A}}}\!=\!0$ so it does not appear.
The performance of this compression channel is numerically calculated in Fig.~\ref{fig:num-QFI}.

\textit{Conclusion.---} {We propose a unified theory, which implies that quantum measurements can be viewed as either information-extracting apparatus as in the standard quantum metrology, or information filters as in the Postselected quantum metrology.} It can be employed
to distribute the optimal measurements through post-selections so
that the cost of the final detections are dramatically reduced, in synergy with recent efforts on distributed quantum sensing
(see e.g.~\citep{malitesta2023distributed,malia2022distributed,zhang2021distributed,ge2018distributed}). As a result, we anticipate our results will find applications
in quantum sensing technologies, such as \textcolor{black}{optical
imaging and interferometry~\citep{demkowicz-dobrzanski2015quantum,tsang2016quantum,zhou2019quantumlimited},
magnetometry~\citep{zhu2022sunlightdriven}, frequency estimation~\citep{dutta2020asingle},
etc.} Many problems are open for future exploration, including the
compression of mixed states, multi-parameter states~\citep{liu2019quantum,szczykulska2016multiparameter},
multipartite-entangled states, etc.

\textit{Acknowledgement.---} The author is grateful to Andrew Jordan and Yiyu Zhou for useful feedback on the manuscript. This work was funded by the Wallenberg
Initiative on Networks and Quantum Information (WINQ) program. 

\let\oldaddcontentsline\addcontentsline     
\renewcommand{\addcontentsline}[3]{}         

\bibliographystyle{apsrev4-1}
\bibliography{Postselected-Many-body-Metrology}

\clearpage\newpage\setcounter{equation}{0} \setcounter{section}{0}
\setcounter{subsection}{0} 
\global\long\def\theequation{S\arabic{equation}}%
\onecolumngrid \setcounter{enumiv}{0} 

\setcounter{equation}{0} \setcounter{section}{0} \setcounter{subsection}{0} \renewcommand{\theequation}{S\arabic{equation}} \onecolumngrid \setcounter{enumiv}{0} \setcounter{figure}{0} \renewcommand{\thefigure}{S\arabic{figure}}
\begin{center}
\textbf{\large{}Supplemental Material}{\large\par}
\par\end{center}

In this Supplemental Material, we present the expression of the quantum Fisher information (QFI) after the post-selection measurement, the saturation conditions for the various bounds of the QFI, proof of Theorem~\ref{thm: LCC-structure} in the main text, calculations on the sensitivity of the Postselected state, weak value amplifications and the example of restricted compression, and expressions of $\mathinner {|{\partial _{x}^{\perp }\psi _{x}}\rangle }\mathinner {\langle {\partial _{x}^{\perp }\psi _{x}}|}$ and $\mathinner {|{\partial _{x}^{\perp }\psi _{x}}\rangle }\mathinner {\langle {\psi _{x}}|}$ in terms of $\rho _{x}$.

\let\addcontentsline\oldaddcontentsline     

\tableofcontents{}

\addtocontents{toc}{\protect\thispagestyle{empty}}
\pagenumbering{gobble}
\section{Analytic expressions of the QFIs }

We define the SLD associated with the state $\sigma_{x|\omega}$
\begin{equation}
\frac{1}{2}\left(L_{x|\omega}^{\text{}}\sigma_{x|\omega}^{\text{}}+\sigma_{x|\omega}^{\text{}}L_{x|\omega}^{\text{}}\right)=\partial_{x}\sigma_{x|\omega}^{\text{}}.\label{eq:L-x-omg-def}
\end{equation}
It is straightforward to check that~\citep{combes2014quantum}
\begin{equation}
L_{x}^{\text{SA}}=\sum_{\omega}[\partial_{x}\ln p(\omega|x)\mathbb{I}^{\text{}}+L_{x|\omega}^{\text{}}]\otimes\ket{\pi_{\omega}^{A}}\bra{\pi_{\omega}^{A}}
\end{equation}
satisfy 
\begin{equation}
\frac{1}{2}\left(L_{x}^{\text{SA}}\sigma_{x}^{\text{SA}}+\sigma_{x}^{\text{SA}}L_{x}^{\text{SA}}\right)=\partial_x\sigma_{x}^{\text{SA}},
\end{equation}
and the QFI associated with $\sigma_{x}^{\text{SA}}$ is 
\begin{equation}
I(\sigma_{x}^{\text{SA}})=\text{Tr}\left[\sigma_{x}^{\text{SA}}\left(L_{x}^{\text{SA}}\right)^{2}\right]=\sum_{\omega}I_{\omega}(\sigma_{x}^{\text{SA}}),
\end{equation}
where 
\begin{equation}
I_{\omega}(\sigma_{x}^{\text{SA}})\equiv I_{\text{\ensuremath{\omega}}}^{\text{cl}}\left(p(\omega\big|x)\right)+p(\omega\big|x)I(\sigma_{x|\omega}^{\text{}}),
\end{equation}
and $I_{\omega}^{\text{cl}}\left(p(\omega\big|x)\right)\equiv\left[\partial_{x}p(\omega\big|x)\right]^{2}/p(\omega\big|x)$.
We denote 
\begin{equation}
\ket{\psi_{x|\omega}}\equiv\frac{K_{\omega}\ket{\psi_{x}}}{\sqrt{p(\omega|x)}}.
\end{equation}
It is straightforward to calculate 
\begin{equation}
\ket{\partial_{x}\psi_{x|\omega}}=\frac{1}{\sqrt{p(\omega|x)}}K_{\omega}\ket{\partial_{x}\psi_{x}}-\frac{\partial_{x}p(\omega|x)}{2p^{3/2}(\omega|x)}K_{\omega}\ket{\psi_{x}},
\end{equation}
\begin{equation}
\braket{\partial_{x}\psi_{x|\omega}\big|\partial_{x}\psi_{x|\omega}}=\frac{\braket{\partial_{x}\psi_{x}\big|K_{\omega}^{\dagger}K_{\omega}\big|\partial_{x}\psi_{x}}}{p(\omega|x)}-\frac{\partial_{x}p(\omega|x)}{p^{2}(\omega|x)}\text{Re}\braket{\partial_{x}\psi_{x}\big|K_{\omega}^{\dagger}K_{\omega}\big|\psi_{x}}+\frac{[\partial_{x}p(\omega|x)]^{2}}{4p^{2}(\omega|x)},
\end{equation}
\begin{equation}
\big|\braket{\partial_{x}\psi_{x|\omega}\big|\psi_{x|\omega}}\big|^{2}=\frac{\big|\braket{\partial_{x}\psi_{x}\big|K_{\omega}^{\dagger}K_{\omega}\big|\psi_{x}}\big|^{2}-\text{Re}\braket{\partial_{x}\psi_{x}\big|K_{\omega}^{\dagger}K_{\omega}\big|\psi_{x}}\partial_{x}p(\omega|x)}{p^{2}(\omega|x)}+\frac{[\partial_{x}p(\omega|x)]^{2}}{4p^{2}(\omega|x)}.
\end{equation}
Subtracting above two equations yields

\begin{equation}
I(\sigma_{x|\omega})=4\left(\frac{\braket{\partial_{x}\psi_{x}\big|E_{\omega}\big|\partial_{x}\psi_{x}}}{\braket{\psi_{x}\big|E_{\omega}\big|\psi_{x}}}-\frac{\big|\braket{\partial_{x}\psi_{x}\big|E_{\omega}\big|\psi_{x}}\big|^{2}}{\braket{\psi_{x}\big|E_{\omega}\big|\psi_{x}}^{2}}\right).
\end{equation}
For later mathematical convenience, let us express all the quantities
in terms of $\ket{\psi_{x}}$ and its orthogonal vector $\ket{\partial_{x}^{\perp}\psi_{x}}$.
It can be readily checked that
\begin{align}
\braket{\partial_{x}\psi_{x}\big|E_{\omega}\big|\partial_{x}\psi_{x}} & =\braket{\partial_{x}^{\perp}\psi_{x}\big|E_{\omega}\big|\partial_{x}^{\perp}\psi_{x}}+(\braket{\partial_{x}^{\perp}\psi_{x}\big|E_{\omega}\big|\psi_{x}}\braket{\psi_{x}\big|\partial_{x}\psi_{x}}+\text{c.c})\nonumber \\
 & +\big|\braket{\partial_{x}\psi_{x}\big|\psi_{x}}\big|^{2}\braket{\psi_{x}\big|E_{\omega}\big|\psi_{x}},\label{eq:dpsi-E-dpsi}
\end{align}
\begin{equation}
\braket{\partial_{x}\psi_{x}\big|E_{\omega}\big|\psi_{x}}=\braket{\partial_{x}^{\perp}\psi_{x}\big|E_{\omega}\big|\psi_{x}}+\braket{\partial_{x}\psi_{x}\big|\psi_{x}}\braket{\psi_{x}\big|E_{\omega}\big|\psi_{x}},\label{eq:dpsi-E-psi}
\end{equation}
\begin{equation}
\big|\braket{\partial_{x}\psi_{x}\big|E_{\omega}\big|\psi_{x}}\big|^{2}=\big|\braket{\partial_{x}^{\perp}\psi_{x}\big|E_{\omega}\big|\psi_{x}}\big|^{2}+(\braket{\partial_{x}^{\perp}\psi_{x}\big|E_{\omega}\big|\psi_{x}}\braket{\psi_{x}\big|\partial_{x}\psi_{x}}\braket{\psi_{x}\big|E_{\omega}\big|\psi_{x}}+\text{c.c})+\big|\braket{\partial_{x}\psi_{x}\big|\psi_{x}}\big|^{2}\braket{\psi_{x}\big|E_{\omega}\big|\psi_{x}}^{2}.
\end{equation}
As a result, we find
\begin{equation}
I_{\omega}^{\text{cl}}\left(p(\omega\big|x)\right)=\frac{[\text{Tr}(\partial_{x}\rho_{x}E_{\omega})]^{2}}{\text{Tr}(\rho_{x}E_{\omega})}=\frac{4\left(\text{Re}\braket{\partial_{x}\psi_{x}\big|E_{\omega}\big|\psi_{x}}\right)^{2}}{\braket{\psi_{x}\big|E_{\omega}\big|\psi_{x}}}=\frac{4\left(\text{Re}\braket{\partial_{x}^{\perp}\psi_{x}\big|E_{\omega}\big|\psi_{x}}\right)^{2}}{\braket{\psi_{x}\big|E_{\omega}\big|\psi_{x}}},\label{eq:I-cl-omg}
\end{equation}
\begin{equation}
I(\sigma_{x|\omega}^{\text{}})=\frac{4\left(\braket{\partial_{x}^{\perp}\psi_{x}\big|E_{\omega}\big|\partial_{x}^{\perp}\psi_{x}}\braket{\psi_{x}\big|E_{\omega}\big|\psi_{x}}-\big|\braket{\partial_{x}^{\perp}\psi_{x}\big|E_{\omega}\big|\psi_{x}}\big|^{2}\right)}{\braket{\psi_{x}\big|E_{\omega}\big|\psi_{x}}^{2}}.\label{eq:I-sigma-given-omg}
\end{equation}
Therefore
\begin{equation}
I_{\omega}(\sigma_{x}^{\text{SA}})=I_{\omega}^{\text{cl}}\left(p(\omega\big|x)\right)+p(\omega\big|x)I(\sigma_{x|\omega}^{\text{S}})=I_{\omega}(\rho_{x})-\frac{4\left(\text{Im}\braket{\partial_{x}^{\perp}\psi_{x}\big|E_{\omega}\big|\psi_{x}}\right)^{2}}{\braket{\psi_{x}\big|E_{\omega}\big|\psi_{x}}},\label{eq:I-sigma-omg-SA}
\end{equation}
where 
\begin{equation}
I_{\omega}(\rho_{x}^{\text{}})\equiv4\braket{\partial_{x}^{\perp}\psi_{x}\big|E_{\omega}\big|\partial_{x}^{\perp}\psi_{x}}.
\end{equation}

\section{\label{sec:bounds}Low and upper bounds of $I_{\omega}$ and $I_{\omega}^{\text{cl}}\left(p(\omega\big|x)\right)$}

Let us note the QFIs must be non-negative. Secondly, according to
Eq.~(\ref{eq:I-sigma-omg-SA}), it is clear that 
\begin{equation}
I_{\omega}(\sigma_{x}^{\text{SA}})\leq I_{\omega}(\rho_{x}^{\text{}}).\label{ineq:I-sig-rho-omg}
\end{equation}
Summing over all the measurement outcome leads to $I(\sigma_{x}^{\text{SA}})\leq I(\rho_{x}^{\text{}})$.
Clearly, given Eq.~(\ref{eq:I-omg}) in the main text, we also have
\begin{equation}
p(\omega\big|x)I(\sigma_{x|\omega}^{\text{}})\leq I_{\omega}(\rho_{x}^{\text{}}),\label{ineq:ave-QFI}
\end{equation}
and 
\begin{equation}
I_{\text{\ensuremath{\omega}}}^{\text{cl}}\left(p(\omega\big|x)\right)\leq I_{\omega}(\rho_{x}^{\text{}}).\label{ineq:I-cl}
\end{equation}
In standard metrology, only the post-selection measurement statistics
matters as the post-measurement states $\sigma_{x|\omega}$ will be
discarded. That is, only $I_{\omega}^{\text{cl}}\left(p(\omega\big|x)\right)$
matters. We can calculate 
\begin{equation}
I_{\omega}^{\text{cl}}\left(p(\omega\big|x)\right)=\frac{4\left(\text{Re}\braket{\partial_{x}^{\perp}\psi_{x}\big|E_{\omega}\big|\psi_{x}}\right)^{2}}{\braket{\psi_{x}\big|E_{\omega}\big|\psi_{x}}}\leq\frac{4\big|\braket{\partial_{x}^{\perp}\psi_{x}\big|E_{\omega}\big|\psi_{x}}\big|^{2}}{\braket{\psi_{x}\big|E_{\omega}\big|\psi_{x}}}\leq I_{\omega}(\rho_{x}^{\text{}}).\label{ineq:I-cl-rho-omg}
\end{equation}
Summing over all the measurement outcome leads to $\sum_{\omega}I_{\omega}^{\text{cl}}\left(p(\omega\big|x)\right)\leq I(\rho_{x})$.

As one can see from Eqs.~(\ref{eq:I-cl-omg},~\ref{eq:I-sigma-given-omg},~\ref{eq:I-sigma-omg-SA}),
the expressions of the QFI all involve the probability of the corresponding
measurement outcome $\braket{\psi_{x}\big|E_{\omega}\big|\psi_{x}}$
in the denominator, one should distinguish the case where $\braket{\psi_{x}\big|E_{\omega}\big|\psi_{x}}\ne0$
and the case where $\braket{\psi_{x}\big|E_{\omega}\big|\psi_{x}}=0$,
which was known as the regular and null POVM operator, respectively,
according to Ref.~\citep{yang2019optimal}. Equivalently, one should
distinguish whether$\sqrt{E_{\omega}}\ket{\psi_{x}}$ vanishes or
not . 

Finally, let us note a trivial case where $\sqrt{E_{\omega}}\ket{\partial_{x}^{\perp}\psi_{x}}=0$.
In this case $I_{\omega}(\rho_{x}^{\text{}})=0$, which dictates $I_{\omega}^{\text{cl}}\left(p(\omega\big|x)\right)=I(\sigma_{x|\omega}^{\text{}})=0$,
according to the non-negativity of the QFI. We shall exclude this
trivial case in the following discussion and the main text. 

\subsection{Regular POVM operator with $\sqrt{E_{\omega}}\ket{\psi_{x}}\protect\neq0$
and $\sqrt{E_{\omega}}\ket{\partial_{x}^{\perp}\psi_{x}}\protect\neq0$}

In this case, If we perform a spectral decomposition of $E_{\omega}=\sum_{\mu}\lambda_{\omega\mu}\ket{e_{\omega\mu}}\bra{e_{\omega\mu}}$,
the it indicates at there exists at least one vector such that $\braket{e_{\omega\mu}\big|\psi_{x}}\neq0$.
According to Eq. ~(\ref{eq:I-sigma-omg-SA}), the inequality~(\ref{ineq:I-sig-rho-omg})
is saturated if Eq.~(\ref{Teq:Im-perp-E}) in the main text is satisfied. 

Next, according to Eq.~(\ref{eq:I-sigma-given-omg}), $I(\sigma_{x|\omega}^{\text{}})$
vanishes if Eq.~(\ref{Teq:prop-cond-c}) is satisfied. Furthermore,
according to Eq.~(\ref{eq:I-cl-omg}), $I_{\text{\ensuremath{\omega}}}^{\text{cl}}\left(p(\omega\big|x)\right)$
vanishes if Eq.~(\ref{Teq:Re-vanish}) in the main text is satisfied. 

To saturate Eq.~(\ref{ineq:ave-QFI}), we would like to saturate
Eq.~(\ref{ineq:I-sig-rho-omg}) but also ensure $I_{\text{\ensuremath{\omega}}}^{\text{cl}}\left(p(\omega\big|x)\right)$
vanishes at the same time. This leads to Eq.~(\ref{Teq:perp-cond})
in the main text. In this case, the measurement statistics contains
no information, but the maximum amount of information all goes into
the selective post-selection state $\sigma_{x|\omega}$. 

Finally, the saturation of Eq.~(\ref{ineq:I-cl-rho-omg}) requires,
in addition to Eq.~(\ref{Teq:Im-perp-E}) in the main text, 
\begin{equation}
\sqrt{E_{\omega}}\ket{\partial_{x}^{\perp}\psi_{x}}\propto\sqrt{E_{\omega}}\ket{\psi_{x}}.
\end{equation}
These two condition leads to Eq.~(\ref{Teq:prop-cond-r}) in the
main text, which was addressed in the paper by Braunstein and Caves~\cite{braunstein1994statistical}
and more recently in Ref.~\cite{yang2019optimal} and Ref.~\cite{len2022quantum}. Indeed, this condition includes
Eq.~(\ref{Teq:prop-cond-c}) in the main text as a special case and
therefore $I(\sigma_{x|\omega}^{\text{S}})=0$. That is, no information
is left after the optimal measurements in standard metrology, which
physically makes sense!

\subsection{Null POVM operator with $\sqrt{E_{\omega}}\ket{\psi_{x}}=0$ and
$\sqrt{E_{\omega}}\ket{\partial_{x}^{\perp}\psi_{x}}\protect\neq0$}

In this case, we know 
\begin{equation}
\braket{e_{\omega\mu}\big|\psi_{x}}=0,\,\forall\mu,
\end{equation}
that is, $\ket{e_{\omega\mu}}$ lies in the kernel of the density
matrix $\rho_{x}^{\text{}}$. 
Using L'hospital rule, Ref.~\citep{yang2019optimal} prove that
\begin{equation}
\lim_{y\to x}I_{\omega}^{\text{cl}}\left(p(\omega\big|y)\right)=I_{\omega}(\rho_{x}^{\text{}}).
\end{equation}
for single-parameter estimation.
This condition implies that when our prior knowledge of the estimation
parameter denoted $x_{*}$ is very close to $x$, the channel $E_{\omega}(x_{*})$
leaves almost no information in the selective post-measurement state
$\sigma_{x|\omega}$.\textcolor{blue}{{} }\textcolor{black}{Indeed,
using L'hospital rule, one can further confirm that $I(\sigma_{x|\omega}^{\text{}})$
vanishes if $E_{\omega}$ is a null POVM operator.}

\section{Proof of Theorem~\ref{thm: LCC-structure}}

As mentioned in Sec.~\ref{sec:bounds}, measurement outcomes with
$\sqrt{E_{\omega}}\ket{\partial_{x}^{\perp}\psi_{x}}=0$ should not
be included in the retained set of post-measurement states as it contains
no information. Therefore, for $\omega\in\checkmark$,
\begin{equation}
\sqrt{E_{\omega}}\ket{\partial_{x}^{\perp}\psi_{x}}\neq0.\label{eq:sqrtE-dperp}
\end{equation}
We denote the orthonormal vectors orthogonal to $\ket{\psi_{x}}$
as $\ket{\psi_{x}^{\perp(n)}}$, where $n=1,\,2,\cdots,\,=\dim\mathcal{H}-1$.
We further choose $\ket{\psi_{x}^{\perp(1)}}$ or $\ket{\psi_{x}^{\perp}}$
as the normalized vector parallel to $\ket{\partial_{x}^{\perp}\psi_{x}}$,
\begin{equation}
\ket{\psi_{x}^{\perp(1)}}\equiv\ket{\psi_{x}^{\perp}}=\frac{\ket{\partial_{x}^{\perp}\psi_{x}}}{\sqrt{\braket{\partial_{x}^{\perp}\psi_{x}\big|\partial_{x}^{\perp}\psi_{x}}}}=\frac{\ket{\partial_{x}^{\perp}\psi_{x}}}{\sqrt{I(\rho_{x})}}.
\end{equation}
One can always decompose $E_{\omega}$ in this basis and write
\begin{equation}
E_{\omega}=q_{\omega}\ket{\psi_{x}^{\perp(1)}}\bra{\psi_{x}^{\perp(1)}}+\Lambda_{\omega},
\end{equation}
where $\Lambda_{\omega}$ is an operator satisfying
\begin{align}
\Lambda_{\omega} & \equiv c_{0}\ket{\psi_{x}}\bra{\psi_{x}}+\sum_{m=2}^{\dim\mathcal{H}-1}c_{m}\ket{\psi_{x}^{\perp(1)}}\bra{\psi_{x}^{\perp(1)}}\nonumber \\
 & +\sum_{m=1}^{\dim\mathcal{H}-1}(c_{0m}\ket{\psi_{x}}\bra{\psi_{x}^{\perp(m)}}+\text{h.c.})+\sum_{m,\,n=1,\,m\neq n}^{\dim\mathcal{H}-1}(c_{mn}\ket{\psi_{x}^{\perp(m)}}\bra{\psi_{x}^{\perp(n)}}+\text{h.c.}).
\end{align}
Apparently, $\Lambda_{\omega}$ satisfies 
\begin{equation}
\braket{\partial_{x}^{\perp}\psi_{x}\big|\Lambda_{\omega}\big|\partial_{x}^{\perp}\psi_{x}}=0.
\end{equation}
Eq.~(\ref{eq:sqrtE-dperp}) implies that 
\begin{equation}
q_{\omega}\neq0.
\end{equation}
Eq.~(\ref{eq:E-omg-cross}) in the main text is equivalent to 
\[
I(\rho_{x}^{\text{}})=\braket{\partial_{x}^{\perp}\psi_{x}\big|\partial_{x}^{\perp}\psi_{x}}=\sum_{\omega\in\checkmark}\braket{\partial_{x}^{\perp}\psi_{x}\big|E_{\omega}\big|\partial_{x}^{\perp}\psi_{x}}.
\]
That is, the Postselected states contain all the QFIs. This leads
to
\begin{equation}
\sum_{\omega\in\checkmark}q_{\omega}=1.
\end{equation}
Since
\begin{equation}
p(\omega|x)=\braket{\psi_{x}\big|E_{\omega}|\psi_{x}}=\braket{\psi_{x}\big|\Lambda_{\omega}\big|\psi_{x}}=\lambda_{\omega},
\end{equation}
to guarantee the LCC is efficient, i.e., we require
\begin{equation}
1<\frac{1}{\sum_{\omega\in\checkmark}\lambda_{\omega}}<\infty.
\end{equation}
That is, $\lambda_{\omega}$ cannot be exactly zero or one. 

Finally, Eq.~(\ref{eq:E-checkmark-general}), i.e., the condition
that the measurement statistics associated with the retained states
carry no information leads to
\begin{equation}
\braket{\partial_{x}^{\perp}\psi_{x}\big|\Lambda_{\omega}\big|\psi_{x}}=0.
\end{equation}

\section{Sensitivity of the Postselected state}

The Postselected state is defined as 
\begin{equation}
\ket{\psi_{x|\omega}}=\frac{K_{\omega}\ket{\psi_{x}}}{\sqrt{\braket{\psi_{x}\big|E_{\omega}\big|\psi_{x}}}}=\frac{K_{\omega}\ket{\psi_{x}}}{\sqrt{\lambda_{\omega}}},
\end{equation}
Furthermore, one can calculate 
\begin{equation}
\ket{\partial_{x}\psi_{x|\omega}}=\frac{K_{\omega}\ket{\partial_{x}\psi_{x}}}{\sqrt{\braket{\psi_{x}\big|E_{\omega}\big|\psi_{x}}}}-\frac{K_{\omega}\ket{\psi_{x}}\text{Re}\braket{\partial_{x}\psi_{x}\big|E_{\omega}\big|\psi_{x}}}{\braket{\psi_{x}\big|E_{\omega}\big|\psi_{x}}^{3/2}}.
\end{equation}
Clearly, according to Eq.~(\ref{eq:dpsi-E-psi}), one can readily
conclude $\text{Re}\braket{\partial_{x}\psi_{x}\big|E_{\omega}\big|\psi_{x}}=0$,
which leads to 
\begin{equation}
\ket{\partial_{x}\psi_{x|\omega}}=\frac{K_{\omega}\ket{\partial_{x}\psi_{x}}}{\sqrt{\lambda_{\omega}}}.
\end{equation}
If we take $\Lambda_{\omega}=\lambda_{\omega}\ket{\psi_{x}}\bra{\psi_{x}}$
and $K_{\omega}=\sqrt{E_{\omega}}=\sqrt{q_{\omega}}\rho_{x}^{\perp}+\sqrt{\lambda_{\omega}}\ket{\psi_{x}}\bra{\psi_{x}}$.
Clearly, the post measurement state becomes 
\begin{equation}
\ket{\psi_{x|\omega}}=\ket{\psi_{x}}.
\end{equation}
 Furthermore, we find 
\begin{equation}
\ket{\partial_{x}\psi_{x|\omega}}=\ket{\partial_{x}\psi_{x}}+\left(\sqrt{\frac{q_{\omega}}{\lambda_{\omega}}}-1\right)\sqrt{g(\rho_{x})}\ket{\psi_{x}^{\perp}}.
\end{equation}

\section{Calculation details on weak-value amplification}

More specifically, we find 
\begin{equation}
\ket{\psi_{x}}=\cos\left(\frac{\theta}{2}\right)\ket{0}\otimes\int du\varphi_{0}(u-x)\ket{u}+\sin\left(\frac{\theta}{2}\right)\ket{1}\otimes\int du\varphi_{0}(u+x)\ket{u}.
\end{equation}
The QFI associated with $\rho_{x}$ is 
\begin{equation}
I(\rho_{x})=4\text{Var}\left(\sigma_{z}\otimes P_{u}\right)_{\ket{\psi_{0}}}=\frac{1}{\sigma^{2}}.
\end{equation}
It can be readily calculated that 
\begin{equation}
\ket{\partial_{x}\psi_{x}}=\frac{1}{2\sigma^{2}}\left[\cos\left(\frac{\theta}{2}\right)\ket{0}\otimes\int du(u-x)\varphi_{0}(u-x)\ket{u}-\sin\left(\frac{\theta}{2}\right)\ket{1}\otimes\int du(u+x)\varphi_{0}(u+x)\ket{u}\right].
\end{equation}
\[
\]
Apparently$\braket{\psi_{x}\big|\partial_{x}\psi_{x}}=0$ and therefore
$\ket{\partial_{x}^{\perp}\psi_{x}}=\ket{\partial_{x}\psi_{x}}$.
In the limit $x\to0$, we find 
\begin{equation}
\ket{\psi_{x=0}}=\left[\cos\left(\frac{\theta}{2}\right)\ket{0}+\sin\left(\frac{\theta}{2}\right)\ket{1}\right]\otimes\int du\varphi_{0}(u)\ket{u},
\end{equation}
\begin{equation}
\ket{\partial_{x}\psi_{x}}|_{x=0}=\frac{1}{2\sigma^{2}}\left[\cos\left(\frac{\theta}{2}\right)\ket{0}-\sin\left(\frac{\theta}{2}\right)\ket{1}\right]\otimes\int duu\varphi_{0}(u)\ket{u}.
\end{equation}
In terms of normalized first-order Hermite-Gaussian functions, where
$\varphi_{1}(u)=\frac{u}{\sigma}\varphi_{0}(u)$, we can rewrite the
above expressions as 
\begin{equation}
\ket{\psi_{x=0}}=\ket{\phi_{\theta}}\otimes\ket{\varphi_{0}},
\end{equation}
\begin{equation}
\ket{\partial_{x}\psi_{x}}|_{x=0}=\frac{1}{2\sigma}\ket{\phi_{-\theta}}\otimes\ket{\varphi_{1}}.
\end{equation}
According to Theorem~\ref{thm: LCC-structure}, we find 
\begin{equation}
E_{\checkmark}=\Pi_{-\theta}\otimes\Pi_{\varphi_{1}}+\Lambda_{\checkmark},
\end{equation}
where $\Lambda_{\checkmark}$ satisfy Eq.~(\ref{eq:Lambda-Condition})
in the main text and $\Pi_{\theta}\equiv\ket{\phi_{\theta}}\bra{\phi_{\theta}}$
, $\Pi_{\varphi_{n}}\equiv\ket{\varphi_{n}}\bra{\varphi_{n}}$. It
is straightforward to verify that 
\begin{equation}
\Lambda_{\checkmark}=\Pi_{-\theta}\otimes\sum_{n\neq1}\Pi_{\varphi_{n}}
\end{equation}
satisfies Eq.~(\ref{eq:Lambda-Condition}). Thus if we construct
\begin{equation}
E_{\checkmark}=\Pi_{-\theta},
\end{equation}
then 
\begin{equation}
\lambda_{\checkmark}=\text{Tr}(\Pi_{-\theta}\Pi_{\theta})=\big|\braket{\phi_{-\theta}\big|\phi_{\theta}}\big|^{2}=\cos^{2}\theta.
\end{equation}
In the case of $\theta=\pi/2$, we choose 
\begin{equation}
\Lambda_{\checkmark}=\Pi_{-\theta}\otimes\sum_{n\neq1}\Pi_{\varphi_{n}}+\varepsilon\Pi_{-\theta}^{\perp}\otimes\mathbb{I}
\end{equation}
which again satisfies Eq.~(\ref{eq:Lambda-Condition}). Finally,
one can also choose 
\begin{equation}
\Lambda_{\checkmark}=\Pi_{-\theta}^{\perp}\otimes\Pi_{\varphi_{1}}+\varepsilon\mathbb{I}\otimes\Pi_{\varphi_{0}},
\end{equation}
so that 
\begin{equation}
E_{\checkmark}=\mathbb{I}\otimes\left(\Pi_{\varphi_{1}}+\varepsilon\Pi_{\varphi_{0}}\right).
\end{equation}

\section{Entangled initial state with the non-interacting Hamiltonian}

We consider bipartite systems denoted as $A$ and $B$, respectively,
whose Hamiltonian is given by 
\begin{equation}
\tilde{H}_{\text{AB}}=\tilde{H}_{A}+\tilde{H}_{B}.\label{eq:sum}
\end{equation}
The final state $\ket{\psi_{x}^{\text{AB}}}=e^{-\text{i}Hx}\ket{\psi_{0}^{\text{AB}}}$.
One can always make the following replacement
\begin{equation}
\tilde{H}_{\text{A}}\to H_{\text{A}}=\tilde{H}_{\text{A}}-\text{Tr}_{\text{AB}}(\rho_{0}^{\text{AB}}\tilde{H}_{\text{A}}),\label{eq:HA-shift}
\end{equation}
such that $\text{Tr}_{\text{AB}}(\rho_{0}^{\text{AB}}H_{\text{A}})=0$.
Similarly, we can make $\text{Tr}_{\text{AB}}(\rho_{0}^{\text{AB}}H_{\text{B}})=0$.
Thus $H_{\text{AB}}$ can be always gauged such that 
\begin{equation}
\braket{\psi_{x}^{\text{AB}}\big|H_{\text{AB}}\big|\psi_{x}^{\text{AB}}}=\braket{\psi_{0}^{\text{AB}}\big|H_{\text{AB}}\big|\psi_{0}^{\text{AB}}}=0\label{eq:ave-vanish}
\end{equation}
by subtracting a constant that depends on the initial state. So without
loss of generality we assume $\text{Tr}_{\text{AB}}(\rho_{0}^{\text{AB}}H_{\text{A}})=\text{Tr}_{\text{AB}}(\rho_{0}^{\text{AB}}H_{\text{B}})=0$.
Thus, we obtain
\begin{equation}
\ket{\partial_{x}\psi_{x}^{\text{AB}}}=\ket{\partial_{x}^{\perp}\psi_{x}^{\text{AB}}}=-\text{i}H_{\text{AB}}\ket{\psi_{x}^{\text{AB}}}
\end{equation}
thanks to $\braket{\psi_{x}^{\text{AB}}\big|\partial_{x}\psi_{x}^{\text{AB}}}=0$.
Furthermore, it is straightforward to find 
\begin{equation}
\rho_{x}^{\text{AB}}=U_{x}^{\text{AB}}\rho_{0}^{\text{AB}}U_{x}^{\text{AB}\dagger},\;\rho_{x}^{\perp\text{AB}}=\frac{U_{x}^{\text{AB}}H_{\text{AB}}\rho_{0}^{\text{AB}}H_{\text{AB}}U_{x}^{\text{AB}\dagger}}{\text{Tr}_{\text{AB}}(\rho_{0}^{\text{AB}}H_{\text{AB}}^{2})},\:\mathcal{C}_{x}^{\text{AB}}=-\text{i}U_{x}^{\text{AB}}H_{\text{AB}}\rho_{0}^{\text{AB}}U_{x}^{\text{AB}\dagger}.\label{eq:key-quantities}
\end{equation}
On the other hand, it can be calculated that,
\begin{equation}
\varrho_{x}^{\text{A}}=U_{x}^{\text{A}}\text{Tr}_{\text{B}}\left(U_{x}^{\text{B}}\rho_{0}^{\text{AB}}U_{x}^{\text{B}\dagger}\right)U_{x}^{\text{A}\dagger}.
\end{equation}
Note that the partial trace still satisfies the cyclic property, i.e.,
\begin{equation}
\text{Tr}_{\text{B}}(R^{\text{B}}T^{\text{AB}})=\text{Tr}_{\text{B}}(R^{\text{B}}T^{\text{AB}}),\label{eq:partial-trace-property}
\end{equation}
which can be immediately proved upon decomposing $T^{\text{AB}}$
as $T^{\text{AB}}=\sum_{ij}w_{ij}M_{i}^{\text{A}}\otimes N_{j}^{\text{B}}$.
Thus we find 
\begin{equation}
\varrho_{x}^{\text{A}}=U_{x}^{\text{A}}\varrho_{0}^{\text{A}}U_{x}^{\text{A}\dagger}.
\end{equation}
Similarly, we find 
\begin{align}
\text{Tr}_{\text{B}}\left(U_{x}^{\text{AB}}H_{\text{AB}}\rho_{0}^{\text{AB}}H_{\text{AB}}^{\text{}}U_{x}^{\text{AB}\dagger}\right) & =\text{Tr}_{\text{B}}\left\{ U_{x}^{\text{A}}U_{x}^{\text{B}}\left(H_{\text{A}}\rho_{0}^{\text{AB}}H_{\text{A}}+H_{\text{A}}\rho_{0}^{\text{AB}}H_{\text{B}}+H_{\text{B}}\rho_{0}^{\text{AB}}H_{\text{A}}+H_{\text{B}}\rho_{0}^{\text{AB}}H_{\text{B}}\right)U_{x}^{\text{A}\dagger}U_{x}^{\text{B}\dagger}\right\} \nonumber \\
 & =U_{x}^{\text{A}}\left[H_{\text{A}}\varrho_{0}^{\text{A}}H_{\text{A}}+H_{\text{A}}\text{Tr}_{\text{B}}\left(\rho_{0}^{\text{AB}}H_{\text{B}}\right)+\text{Tr}_{\text{B}}\left(\rho_{0}^{\text{AB}}H_{\text{B}}\right)H_{\text{A}}+\text{Tr}_{\text{B}}(\rho_{0}^{\text{AB}}H_{\text{B}}^{2})\right]U_{x}^{\text{A}\dagger},\label{eq:cumbersome1}
\end{align}
\begin{equation}
\text{Tr}_{\text{AB}}\left(\rho_{0}^{\text{AB}}H_{\text{AB}}^{2}\right)=\text{Tr}_{\text{AB}}\left\{ \rho_{0}^{\text{AB}}\left(H_{\text{A}}^{2}+2H_{\text{A}}\otimes H_{\text{B}}+H_{\text{B}}^{2}\right)\right\} =\text{Tr}_{\text{A}}\left(\varrho_{0}^{\text{A}}H_{\text{A}}^{2}\right)+2\text{Tr}_{\text{AB}}\left(\rho_{0}^{\text{AB}}H_{\text{A}}\otimes H_{\text{B}}\right)+\text{Tr}_{\text{B}}\left(\varrho_{\text{B}}H_{\text{B}}^{2}\right),\label{eq:cumbersome2}
\end{equation}
\begin{align}
\text{Tr}_{\text{B}}\left(U_{x}^{\text{AB}}H_{\text{AB}}\rho_{0}^{\text{AB}}U_{x}^{\text{AB}\dagger}\right) & =\text{Tr}_{\text{B}}\left\{ U_{x}^{\text{A}}U_{x}^{\text{B}}\left(H_{\text{A}}\rho_{0}^{\text{AB}}+H_{\text{B}}\rho_{0}^{\text{AB}}\right)U_{x}^{\text{A}\dagger}U_{x}^{\text{B}\dagger}\right\} \nonumber \\
 & =U_{x}^{\text{A}}\left[H_{\text{A}}\varrho_{0}^{\text{A}}+\text{Tr}_{\text{B}}\left(\rho_{0}^{\text{AB}}H_{\text{B}}\right)\right]U_{x}^{\text{A}\dagger}.\label{eq:cumbersome3}
\end{align}

As of now, Eqs.~(\ref{eq:cumbersome1}-~\ref{eq:cumbersome3}) still
look cumbersome. Now let us make a remarkable construction, which
will simplify them dramatically. We now make use of a crucial observation:
the average energy of these states $\ket{\phi_{k}^{\text{A}}}$ are
the same. This means the after the shift renormalization of $H_{\text{A}}$,
i.e.,
\begin{equation}
\tilde{H}_{\text{A}}\to H_{\text{A}}=\tilde{H}_{\text{A}}-\mathcal{\mathcal{E}},
\end{equation}
we have
\begin{equation}
\braket{\phi_{k}^{\text{A}}\big|H_{\text{A}}\big|\phi_{k}^{\text{A}}}=0.
\end{equation}
By definition, the action of $H_{\text{A}}$ on $\ket{\phi_{k}^{\text{A}}}$
cannot bring it outside the subspace $\mathcal{V}_{k}^{\text{A}}$,
i.e.,
\begin{equation}
\ket{\phi_{k}^{\text{A}}},\quad H_{\text{A}}\ket{\phi_{k}^{\text{A}}}\in\mathcal{V}_{k}^{\text{A}}.
\end{equation}
 As a consequence,
\begin{equation}
H_{\text{A}}\ket{\phi_{k}^{\text{A}}}\perp\ket{\phi_{k}^{\text{A}}},\:\ket{\phi_{k}^{\text{A}}}\perp\ket{\phi_{l}^{\text{A}}},\,H_{\text{A}}\ket{\phi_{k}^{\text{A}}}\perp\ket{\phi_{l}^{\text{A}}},\:H_{\text{A}}\ket{\phi_{k}^{\text{A}}}\perp H_{\text{A}}\ket{\phi_{l}^{\text{A}}}.\label{eq:orthogonality}
\end{equation}
We define 
\begin{align}
\varrho_{0}^{\text{A}} & \equiv\sum_{k}p_{k}\ket{\phi_{k}^{\text{A}}}\bra{\phi_{k}^{\text{A}}},\label{eq:rho0-A}\\
\ket{\psi_{0}^{\text{AB}}} & =\sum_{k}\sqrt{p_{k}}\ket{\phi_{k}^{\text{A}}}\otimes\ket{\varphi_{k}^{\text{B}}}\label{eq:psi0-AB}\\
\ket{\phi_{k}^{\perp\text{A}}} & \equiv\frac{H_{\text{A}}\ket{\phi_{k}^{\text{A}}}}{\sqrt{\braket{\phi_{k}^{\text{A}}\big|H_{\text{A}}^{2}\big|\phi_{k}^{\text{A}}}}},
\end{align}
where $p_{k}\in(0,1]$ and $\ket{\psi_{0}^{\text{AB}}}$ is the purification
of $\rho_{0}^{\text{A}}$. With such a construction of the initial
state, we find 
\begin{equation}
\text{Tr}_{\text{AB}}\left(\rho_{0}^{\text{AB}}H_{\text{A}}\otimes H_{\text{B}}\right)=\langle\psi_{0}^{\text{AB}}\big|H_{\text{A}}\otimes H_{\text{B}}\big|\psi_{0}^{\text{AB}}\rangle=\sum_{kl}\sqrt{p_{k}p_{l}}\braket{\phi_{k}^{\text{A}}\big|H_{\text{A}}\big|\phi_{l}^{\text{A}}}\braket{\phi_{k}^{\text{B}}\big|H_{\text{B}}\big|\phi_{l}^{\text{B}}}=0,\label{eq:Tr-Simplify1}
\end{equation}
thanks to Eq.~(\ref{eq:orthogonality}). Therefore, we see that the
QFI in the state before post-selection is 
\begin{equation}
I(\rho_{x}^{\text{AB}})=4\text{Var}\left(H_{\text{A}}+H_{\text{B}}\right)_{\rho_{0}^{\text{AB}}}=4\left(\delta h_{\text{A}}^{2}+\delta h_{\text{B}}^{2}\right),
\end{equation}
where where 
\begin{align}
\delta h_{\text{A}} & \equiv\sqrt{\text{Var}(H_{\text{A}})_{\varrho_{\text{A}}}}=\sqrt{\sum_{k}p_{k}\braket{\phi_{k}^{\text{A}}\big|H_{\text{A}}^{2}\big|\phi_{k}^{\text{A}}}},\\
\delta h_{\text{A}} & \equiv\sqrt{\text{Var}(H_{\text{B}})_{\varrho_{\text{B}}}}=\sqrt{\sum_{k}p_{k}\braket{\varphi_{k}^{\text{B}}\big|H_{\text{B}}^{2}\big|\varphi_{k}^{\text{B}}}}.
\end{align}

We further define 
\begin{equation}
\tilde{E}_{k}^{\text{A}}\equiv\ket{\phi_{k}^{\perp\text{A}}}\bra{\phi_{k}^{\perp\text{A}}},
\end{equation}
\begin{align}
\tilde{E}_{\checkmark}^{\text{A}} & \equiv\sum_{k}\tilde{E}_{k}^{\text{A}}=\sum_{k}\ket{\phi_{k}^{\perp\text{A}}}\bra{\phi_{k}^{\perp\text{A}}}.\label{eq:E-tild-def}
\end{align}
 Then it straightforward to check 
\begin{align}
\text{Tr}_{\text{A}}\left\{ H_{\text{A}}\text{Tr}_{\text{B}}\left(\rho_{0}^{\text{AB}}H_{\text{B}}\right)\tilde{E}_{k}^{\text{A}}\right\}  & =\text{Tr}_{\text{B}}\left\{ \text{Tr}_{\text{A}}\left(H_{\text{A}}\rho_{0}^{\text{AB}}\tilde{E}_{k}^{\text{A}}\right)H_{\text{B}}\right\} \nonumber \\
 & =\langle\pi_{l}^{\text{B}}\big|\langle\phi_{k}^{\perp\text{A}}\big|H_{\text{A}}\big|\psi_{0}^{\text{AB}}\rangle\langle\psi_{0}^{\text{AB}}\big|\phi_{k}^{\perp\text{A}}\rangle H_{\text{B}}\big|\pi_{l}^{\text{B}}\rangle\nonumber \\
 & =0,\label{eq:Tr-Simplify2}
\end{align}
thanks to $\langle\psi_{0}^{\text{AB}}\big|\phi_{k}^{\perp\text{A}}\rangle=0$,
and
\begin{align}
\text{Tr}_{\text{A}}\left\{ \text{Tr}_{\text{B}}\left(\rho_{0}^{\text{AB}}H_{\text{B}}^{2}\right)\tilde{E}_{k}^{\text{A}}\right\}  & =\text{Tr}_{\text{B}}\left\{ \text{Tr}_{\text{A}}\left(\rho_{0}^{\text{AB}}\tilde{E}_{k}^{\text{A}}\right)H_{\text{B}}^{2}\right\} \nonumber \\
 & =\langle\pi_{l}^{\text{B}}\big|\langle\phi_{k}^{\perp\text{A}}\big|\psi_{0}^{\text{AB}}\rangle\langle\psi_{0}^{\text{AB}}\big|\phi_{k}^{\perp\text{A}}\rangle H_{\text{B}}^{2}\big|\pi_{l}^{\text{B}}\rangle\nonumber \\
 & =0.\label{eq:Tr-Simplify3}
\end{align}
Upon defining 
\begin{align}
\tilde{E}_{k,\,x}^{\text{A}} & \equiv U_{x}^{\text{A}}\ket{\phi_{k}^{\perp\text{A}}}\bra{\phi_{k}^{\perp\text{A}}}U_{x}^{\text{A}\dagger}=U_{x}^{\text{A}}\tilde{E}_{k}^{\text{A}}U_{x}^{\text{A}\dagger},\\
\tilde{E}_{\checkmark,\,x}^{\text{A}} & \equiv U_{x}^{\text{A}}\sum_{k}\ket{\phi_{k}^{\perp\text{A}}}\bra{\phi_{k}^{\perp\text{A}}}U_{x}^{\text{A}\dagger}=U_{x}^{\text{A}}\tilde{E}_{\checkmark}^{\text{A}}U_{x}^{\text{A}\dagger},
\end{align}
and in view of Eqs.~(\ref{eq:Tr-Simplify1},~\ref{eq:Tr-Simplify2},~\ref{eq:Tr-Simplify3}),
we obtain the following results, 
\begin{equation}
\text{Tr}_{\text{A}}\left(\mathcal{C}_{x}^{\text{A}}\tilde{E}_{k,\,x}^{\text{A}}\right)=\text{Tr}_{\text{A}}\left(H_{\text{A}}\varrho_{0}^{\text{A}}\tilde{E}_{k}^{\text{A}}\right)=p_{k}\sqrt{\braket{\phi_{k}^{\text{A}}\big|H_{\text{A}}^{2}\big|\phi_{k}^{\text{A}}}}\braket{\phi_{k}^{\perp\text{A}}\big|\tilde{E}_{k}^{\text{A}}\big|\phi_{k}^{\text{A}}},\label{eq:Tr-simplified-perp}
\end{equation}
\begin{equation}
\text{Tr}_{\text{A}}\left(\varrho_{x}^{\perp\text{A}}\tilde{E}_{\checkmark,x}^{\text{A}}\right)=\frac{\text{Tr}_{\text{A}}\left(\varrho_{0}^{\text{A}}H_{\text{A}}\tilde{E}_{\checkmark}^{\text{A}}H_{\text{A}}\right)}{\text{Tr}_{\text{A}}\left(\varrho_{0}^{\text{A}}H_{\text{A}}^{2}\right)+\text{Tr}_{\text{B}}\left(\varrho_{\text{B}}H_{\text{B}}^{2}\right)}=\frac{\sum_{k}p_{k}\braket{\phi_{k}^{\text{A}}\big|H_{\text{A}}^{2}\big|\phi_{k}^{\text{A}}}\braket{\phi_{k}^{\perp\text{A}}\big|\tilde{E}_{\checkmark}^{\text{A}}\big|\phi_{k}^{\perp\text{A}}}}{\sum_{k}p_{k}\braket{\phi_{k}^{\text{A}}\big|H_{\text{A}}^{2}\big|\phi_{k}^{\text{A}}}+\text{Tr}_{\text{B}}\left(\varrho_{\text{B}}H_{\text{B}}^{2}\right)},\label{eq:Tr-simplified-noloss}
\end{equation}
\begin{equation}
\text{Tr}_{\text{A}}\left(\varrho_{x}^{\text{A}}\tilde{E}_{k,\,x}^{\text{A}}\right)=\text{Tr}_{\text{A}}\left(\varrho_{0}^{\text{A}}\tilde{E}_{k}^{\text{A}}\right)=p_{k}\braket{\phi_{k}^{\text{A}}\big|\tilde{E}_{\checkmark}^{\text{A}}\big|\phi_{k}^{\text{A}}}.
\end{equation}
Remarkably, we see that a similar structure with Theorem~\ref{thm: LCC-general}
in the main text also occurs in Eqs.~(\ref{eq:Tr-simplified-perp},~\ref{eq:Tr-simplified-noloss})
thanks to the factors $\braket{\phi_{k}^{\perp\text{A}}\big|\tilde{E}_{\checkmark}^{\text{A}}\big|\phi_{k}^{\text{A}}}$
and $\braket{\phi_{k}^{\perp\text{A}}\big|\tilde{E}_{\checkmark}^{\text{A}}\big|\phi_{k}^{\perp\text{A}}}$.
Clearly, with the choice of $\tilde{E}_{\checkmark}^{\text{A}}$ given
by Eq.~(\ref{eq:E-tild-def}), 
\begin{equation}
\text{Tr}_{\text{A}}\left(\mathcal{C}_{x}^{\text{A}}\tilde{E}_{k,\,x}^{\text{A}}\right)=0,\,\text{Tr}_{\text{A}}\left(\varrho_{x}^{\text{A}}\tilde{E}_{k,\,x}^{\text{A}}\right)=0,
\end{equation}
 but 
\begin{equation}
\text{Tr}_{\text{A}}\left(\varrho_{x}^{\perp\text{A}}\tilde{E}_{\checkmark,x}^{\text{A}}\right)=1-\frac{\delta h_{\text{B}}^{2}}{\delta h_{\text{A}}^{2}+\delta h_{\text{B}}^{2}}<1,
\end{equation}
The factor $\frac{\delta h_{\text{B}}^{2}}{\delta h_{\text{A}}^{2}+\delta h_{\text{B}}^{2}}$
characterize the QFI loss in the discarded set. Assuming the energy
fluctuation of the subsystem A dominates over the one of the subsystem
B, the loss is of the order $\sim(\delta h_{\text{B}}/\delta h_{\text{A}})^{2}$.

Now one can construct the compression channel with a very similar
structure to the one in Theorem~\ref{thm: LCC-structure} in the
main text:
\begin{equation}
E_{\omega,\,x}^{A}=\sum_{k}r_{\omega k}\tilde{E}_{k,\,x}^{\text{A}}+\varepsilon U_{x}^{\text{A}}\mathcal{P}_{\text{supp}(\varrho^{\text{A}})}U_{x}^{\text{A}\dagger},\label{eq:E-omg-x}
\end{equation}
where $\sum_{\omega}r_{\omega k}=1$ such that 
\begin{equation}
\sum_{\omega\in\checkmark}E_{\omega,\,x}^{A}=\tilde{E}_{\checkmark,\,x}^{\text{A}}+O(L\varepsilon).
\end{equation}
In the main text, it was mentioned that for the case of restricted post-selection, Theorem~\ref{thm: LCC-general}
becomes 
\begin{equation}
\text{\text{Tr}\ensuremath{\left(\varrho_{x}^{\perp\text{A}}\sum_{\omega\in\checkmark}E_{\omega}^{\text{A}}\right)}}=1,\label{eq:subsystem-LCC-lossless}
\end{equation}
\begin{equation}
\text{Tr}\left(\mathcal{C}_{x}^{\text{A}}E_{\omega}^{\text{A}}\right)=0,\,\omega\in\checkmark,\label{eq:subsystem-LCC-perp}
\end{equation}
with $\text{Tr}\left(\varrho_{x}^{\text{A}}\sum_{\omega\in\checkmark}E_{\omega}^{\text{A}}\right)<1$. 
The figure of merits then can obtained upon inserting
Eq.~(\ref{eq:E-omg-x}) into Eq.~(\ref{eq:subsystem-LCC-lossless}),
Eq.~(\ref{eq:subsystem-LCC-perp}) and $\text{Tr}\left(\varrho_{x}^{\text{A}}\sum_{\omega\in\checkmark}E_{\omega,\,x}^{\text{A}}\right)$.

For the three-qubit example considered in the main text, one can compute
\begin{equation}
\braket{\phi_{k}^{\text{A}}\big|H_{\text{A}}\big|\phi_{k}^{\text{A}}}=\braket{\phi_{2}^{\text{A}}\big|H_{\text{A}}^{2}\big|\phi_{2}^{\text{A}}}=0,\;k=1,\,2,\;\braket{\phi_{1}^{\text{A}}\big|H_{\text{A}}^{2}\big|\phi_{1}^{\text{A}}}=4\omega_{0}^{2},
\end{equation}
\begin{equation}
\braket{\varphi_{1}\big|H_{\text{B}}\big|\varphi_{1}}=-\braket{\varphi_{2}\big|H_{\text{B}}\big|\varphi_{2}}=\cos\theta,\:\braket{\phi_{k}^{\text{A}}\big|H_{\text{B}}^{2}\big|\phi_{k}^{\text{A}}}=\mathbb{I},\:k=1,\,2.
\end{equation}
Therefore, we find 
\begin{equation}
\delta h_{\text{A}}=2\omega_{0}\sqrt{p_{1}},\:\delta h_{\text{B}}=2\Delta\sqrt{p_{1}p_{2}\cos^{2}\theta+\sin^{2}\theta/4}
\end{equation}

\section{\label{sec:expressions}Expression for $\ket{\partial_{x}^{\perp}\psi_{x}}\bra{\partial_{x}^{\perp}\psi_{x}}$
and $\ket{\partial_{x}^{\perp}\psi_{x}}\bra{\psi_{x}}$ in terms of
$\rho_{x}$}

It is straightforward to compute 
\begin{align}
\ket{\partial_{x}^{\perp}\psi_{x}}\bra{\partial_{x}^{\perp}\psi_{x}} & =\left(\ket{\partial_{x}\psi_{x}}-\ket{\psi_{x}}\braket{\psi_{x}\big|\partial_{x}\psi_{x}}\right)\left(\bra{\partial_{x}\psi_{x}}-\bra{\psi_{x}}\braket{\partial_{x}\psi_{x}\big|\psi_{x}}\right)\nonumber \\
 & =\ket{\partial_{x}\psi_{x}}\bra{\partial_{x}\psi_{x}}-\braket{\psi_{x}\big|\partial_{x}\psi_{x}}\ket{\psi_{x}}\bra{\partial_{x}\psi_{x}}-\braket{\partial_{x}\psi_{x}\big|\psi_{x}}\ket{\partial_{x}\psi_{x}}\bra{\psi_{x}}\nonumber \\
 & +\braket{\partial_{x}\psi_{x}\big|\psi_{x}}\braket{\psi_{x}\big|\partial_{x}\psi_{x}}\ket{\psi_{x}}\bra{\psi_{x}}.
\end{align}
On the other hand, 
\begin{align}
\left(\partial_{x}\rho_{x}\right)^{2} & =\left(\ket{\partial_{x}\psi_{x}}\bra{\psi_{x}}+\ket{\psi_{x}}\bra{\partial_{x}\psi_{x}}\right)\left(\ket{\psi_{x}}\bra{\partial_{x}\psi_{x}}+\ket{\partial_{x}\psi_{x}}\bra{\psi_{x}}\right)\nonumber \\
 & =\ket{\partial_{x}\psi_{x}}\bra{\partial_{x}\psi_{x}}+\braket{\partial_{x}\psi_{x}\big|\psi_{x}}\ket{\psi_{x}}\bra{\partial_{x}\psi_{x}}+\braket{\psi_{x}\big|\partial_{x}\psi_{x}}\ket{\partial_{x}\psi_{x}}\bra{\psi_{x}}\nonumber \\
 & +\braket{\partial_{x}\psi_{x}\big|\partial_{x}\psi_{x}}\ket{\psi_{x}}\bra{\psi_{x}}.
\end{align}
Apparently, the above two expressions differ by the last term. Furthermore,
we note 
\begin{align}
\braket{\partial_{x}\psi_{x}\big|\psi_{x}}\braket{\psi_{x}\big|\partial_{x}\psi_{x}} & =\bra{\partial_{x}\psi_{x}}\left(\mathbb{I}-\ket{\psi_{x}^{\perp(1)}}\bra{\psi_{x}^{\perp(1)}}\right)\ket{\partial_{x}\psi_{x}}\nonumber \\
 & =\braket{\partial_{x}\psi_{x}\big|\partial_{x}\psi_{x}}-\braket{\partial_{x}\psi_{x}\big|\psi_{x}^{\perp(1)}}\braket{\psi_{x}^{\perp(1)}\big|\partial_{x}\psi_{x}}\nonumber \\
 & =\braket{\partial_{x}\psi_{x}\big|\partial_{x}\psi_{x}}-g(\rho_{x}).
\end{align}
Thus we conclude 
\begin{equation}
\ket{\partial_{x}^{\perp}\psi_{x}}\bra{\partial_{x}^{\perp}\psi_{x}}=\left(\partial_{x}\rho_{x}\right)^{2}-g(\rho_{x})\rho_{x}.
\end{equation}
The projector along $\ket{\partial_{x}^{\perp}\psi_{x}}$ can then
be expressed as
\begin{equation}
\rho_{x}^{\perp}=\frac{\left(\partial_{x}\rho_{x}\right)^{2}}{g(\rho_{x})}-\rho_{x}.
\end{equation}
Following similar procedure we find, 
\begin{align}
\left(\partial_{x}\rho_{x}\right)\rho_{x} & =\ket{\partial_{x}\psi_{x}}\bra{\psi_{x}}+\ket{\psi_{x}}\braket{\partial_{x}\psi_{x}\big|\psi_{x}}\bra{\psi_{x}},\\
\rho_{x}\left(\partial_{x}\rho_{x}\right) & =\ket{\psi_{x}}\braket{\psi_{x}\big|\partial_{x}\psi_{x}}\bra{\psi_{x}}+\ket{\psi_{x}}\bra{\partial_{x}\psi_{x}}.
\end{align}
Taking the difference between the equations leads to
\begin{equation}
[\partial_{x}\rho_{x},\,\rho_{x}]=\ket{\partial_{x}^{\perp}\psi_{x}}\bra{\psi_{x}}-\ket{\psi_{x}}\bra{\partial_{x}^{\perp}\psi_{x}}.
\end{equation}
Furthermore we find
\begin{align}
 & \ket{\partial_{x}^{\perp}\psi_{x}}\bra{\psi_{x}}+\ket{\psi_{x}}\bra{\partial_{x}^{\perp}\psi_{x}}\nonumber \\
= & \left(\ket{\partial_{x}\psi_{x}}-\ket{\psi_{x}}\braket{\psi_{x}\big|\partial_{x}\psi_{x}}\right)\bra{\psi_{x}}+\ket{\psi_{x}}\left(\bra{\partial_{x}\psi_{x}}-\braket{\partial_{x}\psi_{x}\big|\psi_{x}}\bra{\psi_{x}}\right)\nonumber \\
= & \partial_{x}\rho_{x}.
\end{align}
Thus, we conclude
\begin{equation}
\ket{\partial_{x}^{\perp}\psi_{x}}\bra{\psi_{x}}=\frac{1}{2}\left([\partial_{x}\rho_{x},\,\rho_{x}]+\partial_{x}\rho_{x}\right).
\end{equation}

\end{document}